\documentclass{aa}

\usepackage{graphicx}
\usepackage[dvipsnames]{xcolor}
\usepackage{soul} 
\usepackage[normalem]{ulem}
\usepackage{comment} 
\usepackage{subcaption}

\usepackage{txfonts}
\usepackage{amssymb}
\usepackage{amsmath}

\usepackage[colorlinks=true,
    linkcolor=blue,
    citecolor=blue,
    filecolor=magenta,      
    urlcolor=cyan]{hyperref}

\begin{document} 

   \title{Identification of periodicities with arbitrary shapes in AGN light curves}
   
   \titlerunning{Spotting non-sinusoidal periodicity }
   \authorrunning{L. Bertassi et al.}

   \author{Lorenzo Bertassi\inst{1,2,3}\fnmsep\thanks{l.bertassi@campus.unimib.it} \and
   Maria Charisi\inst{4,5} \and 
   Riccardo Buscicchio\inst{1,2,6} \and
   Fabio Rigamonti\inst{2,3,7} \and 
   Jessie Runnoe\inst{8,9} \and 
   Massimo Dotti \inst{1,2,3}
}

   \institute{
            Universit\`a degli Studi di Milano-Bicocca, Piazza della Scienza 3, 20126 Milano, Italy
        \and
            INFN, Sezione di Milano-Bicocca, Piazza della Scienza 3, I-20126 Milano, Italy
        \and
            INAF - Osservatorio Astronomico di Brera, via Brera 20, I-20121 Milano, Italy
        \and
            Department of Physics and Astronomy, Washington State University, Pullman, WA 99163, USA
        \and
            Institute of Astrophysics, FORTH, GR-71110, Heraklion, Greece
        \and
            Institute for Gravitational Wave Astronomy \& School of Physics and Astronomy, University of Birmingham, Birmingham, B15 2TT, UK
        \and
        Como Lake centre for AstroPhysics (CLAP), DiSAT, Università dell’Insubria, via Valleggio 11, 22100 Como, Italy
        \and 
        Department of Physics \& Astronomy, Vanderbilt University, Nashville, TN 37240, USA
        \and 
        Department of Life and Physical Sciences, Fisk University, Nashville, TN 37208, USA   
    }

   \date{Received XXX; accepted YYY}

\abstract{
Massive black hole binaries are expected to be observable as periodic AGN in time-domain photometric surveys. Periodicities may originate from different physical processes, including the intermittent gas feeding of the black holes caused by the time-varying non-axisymmetric binary potential, the Doppler boosting of the flux emitted by individual accretion discs bound to the orbiting BHs, and the gravitational lensing of the accretion disc of one black hole due to the presence of the other. Only the Doppler boost scenario applied to circular binaries with non-modulated accretion predicts a sinusoidal light curve, while in the general case, binary signals are expected to show more complex periodic patterns. Current searches for massive black hole binaries rely on techniques tailored to quasi-sinusoidal light curves, but fail to identify the more complex periodicities predicted. We present an alternative method that leverages Gaussian processes, making use of a generic periodic kernel flexible enough for the identification of arbitrary periodicities in unevenly sampled light curves with realistic quasar noise. We demonstrate that it outperforms previously proposed strategies in identifying general periodicities by analysing mock light curves with different baselines. Specifically, we find that our analysis can detect non-sinusoidal periodicities (e.g., sawtooth-shaped or symmetric flares) and retrieves a higher fraction of true periodicities when compared to periodogram analysis or Gaussian processes analysis with less flexible periodic kernels. Furthermore, by comparing the retrieved fraction of periodicities between mock Palomar transient factory light curves and mock Legacy Survey of Space and Time light curves, we find that our analysis is most sensitive to the number of observed cycles. 
The application of this analysis has the potential to greatly increase the scientific return of current and upcoming large time-domain photometric surveys.

}


   \maketitle
%
\section{Introduction} \label{sec:introduction}
Since every galaxy is expected to host a massive black hole \citep[see][]{Kormendy2001}, massive black hole binaries (MBHBs) are considered the natural outcome of galaxy mergers
\citep{BBR80}. If the interaction with their local environment is effective in removing orbital energy and angular momentum, MBHBs can reach small enough separations and emit low-frequency gravitational waves. These can be detectable by ongoing pulsar timing array (PTA) campaigns \citep{pta} and by future space-based interferometers \citep[e.g., the Laser Interferometer Space Antenna, LISA;][]{lisa1,lisa2}. These binaries are also expected to emit bright electromagnetic signals that can be used to detect them. Multiple observational features identifying MBHB candidates have been proposed (see \citealt{DSD12, Derosa2019, Bogdanovic2021, DOrazio2023} for an overview, and \citealt{Gaskell88, polar22, dotti23b, Chan25, bertassi2025} for alternative signatures). The electromagnetic identification of MBHBs can complement the interpretation of the (still tentative) detection of a  GW background in PTA data \citep{Antoniadis23, Agazie23, Reardon23, Xu23}, and constrain the highly uncertain rates of MBHB coalescences detectable by LISA \citep[e.g.][]{fiacconi13, delvalle15, souzalima17, bortolas20, bortolas22}.

At small separations, corresponding to MBHB orbital periods of a few years, a promising feature for the identification of active MBHBs is the quasi-periodic modulation of their brightness. 
Depending on the binary intrinsic properties, such modulation can be driven by: \textit{(i)} the time varying non-axisymmetric potential of the binary, prompting intermittent gas feeding from the outer circumbinary disc \citep{MacFadyen2008, HMH08}, \textit{(ii)}
the phase-dependent Doppler boosting of the flux emitted by the individual accretion discs bound to the individual BHs (mini-disks) that move with relativistic velocities \citep{DHS15}, and/or \textit{(iii)} the gravitational lensing of the accretion disc of one black hole due to the presence of the other \citep{doraziolense18, Kelley21}.  Periodicities due to the presence of MBHBs can also be found when both MBHs are inactive, e.g. if a luminous star lies behind an MBHB, it can be gravitationally lensed periodically by the binary \citep[see][]{lensing25}\footnote{The list of physical processes imprinting a periodicity on the observed light curve is not definitely complete, as other processes might contribute to a periodic modulation \citep[see e.g.][]{Chan25}.}.

Hundreds of close MBHB candidates have already been selected due to the seeming periodicity of their light curves in time-domain photometric surveys \citep[see][]{Graham15, Charisi2016, LiuGez+2019, Chen+2020, Li2023, chen24, foustoul25, erosita25}. The natural variability of active galactic nuclei, characterised by red noise, i.e. a noise that shows higher power at long timescale variability \citep[see][]{Ulrich1997}, can mimic periodic light curves over a few (seeming) cycles \citep[see][]{Vaughan16, Witt2022, elbadry2025}. Thus,  it is imperative to quantify the statistical evidence in favour of periodic signals when compared to a red noise model component only. Upcoming time-domain photometric surveys, such as the Legacy Survey of Space and Time \citep[LSST, see][]{LSST} and the Nancy Grace Roman Space Telescope \citep[see][]{haiman23}, have the observational length, depth, and light curve sampling frequency needed to identify close-separation MBHB candidates \citep[see e.g. the discussions in][]{Kelley19, xin21, Kelley21, haiman23}.

All the above-mentioned searches looking for periodic light curves \citep[with the exception of][as discussed in what follows]{foustoul25} are based, at least in the pre-selection of the candidates, on the analysis of Lomb Scargle periodograms \citep[LSP, see][]{Lomb1976, Scargle1982, VanderPlas18}, a generalisation of the Fourier transform power spectrum for unevenly sampled timeseries.\footnote{Uneven sampling and gaps in the observations of individual objects are unavoidable, due to weather interruptions, sources falling below the horizon or very close to the Sun during certain seasons. Space-based observatories achieve more regular coverage, but still experience interruptions due to operational constraints.} 
LSP-based analysis suffers from three main limitations: \textit{(i)}  the general distribution of the periodogram peaks is not known (limiting our ability to estimate the likelihood of a period), \textit{(ii)}
the independence of the specific powers at each frequency bin
is not guaranteed \citep[e.g.][]{Covino2022, Lin2025}, and, most importantly,  $(iii)$ non-sinusoidal periodic signals will spread power over multiple frequencies, making the identification of a single statistically significant peak in the power spectrum more challenging. While points \textit{(i)} and \textit{(ii)} apply only to unevenly sampled light curves (for which the LSP cannot be constructed directly through a Fourier transform), point $(iii)$ applies to the unrealistic scenario of evenly spaced observations, too.  
Indeed \cite{Lin2025} showed that, in the presence of realistic quasar noise, LSP-based searches identify only $\lesssim 40 \%$ of sinusoidal signals ---for which they are best suited---, and $\lesssim 10 \%$ of more complex (sawtooth-shaped) periodic signals. 
This is particularly relevant, since only Doppler boosting of circular binaries with non-modulated accretion can produce sinusoidal light curves.
All other physical mechanisms of periodicity instead show more complex and irregular profiles \citep[see][]{Hu2020,duffell20, Zrake_2021,Westernacher_Schneider22, Cocchiararo24}.

A few periodic light curve searches have been published that are not based on the LSP (see \citealt{covino20, Zhu_2020,  rigamonti25} for applications to single objects and \citealt{foustoul25} for larger searches). Therein, light curves are modelled directly in the time-domain through Gaussian Processes \citep[GPs, see][]{Rasmussen2006, aigrain2022}, for which a likelihood, and hence the evidence of a periodic model, is well defined. However, the reliability of the evidence depends on the validity of the assumed noise model. 
The GP kernel associated with a periodic model is often described by a simple cosine function. As detailed in Section~\ref{sec:algorithm}, this prevents a GP model from modelling sudden jumps in the light curves over timescales shorter than the period, $P$. We note that the GPs have already been applied in AGN studies, for example, in searching for flares \citep[see][]{McLaughlin2024}, and intrinsic variability studies \citep[see][]{Zhang2023, Zhang2024, Weixiang2026}

In this work, we present an alternative analysis based on light-curve modelling through GPs, assuming a more flexible family of periodic kernels, as proposed by \cite{durrande2016}. 
We test them on realistic (periodic and non-periodic) AGN light curves, including the effect of red noise. We show that GPs, once coupled with a robust Bayesian treatment, allow for the detection of complex and irregularly shaped light curves. 
We note that another appealing approach to search for periodicities in astrophysical light curves is the use of deep learning models and neural networks \citep[see][for two examples]{miller2024, fernandes2025}, and indeed such techniques are being applied to real and mock AGN periodic light curves \citep[see][]{Kovacevic2023}. Different from such methods, the GP analysis we are proposing can directly link the kernel parameters to the properties of the light curve, such as the amplitude, the shape and the period and can estimate, as a natural byproduct, the parameters of the intrinsic variability.
In Section~\ref{sec:algorithm}, we discuss the periodicity detection algorithm and compare various GP kernels. 
In Section~\ref{sec:results}, we identify a detection statistic and an optimal threshold in the identification of periodicities. We discuss the fraction of retrieved periodicities and the goodness of the parameter estimation. Finally, we discuss which parameters affect the retrieved fractions the most and compare our findings with alternative methods in the literature. 
In Section~\ref{sec:conclusions} we summarise our conclusions.

\section{Search algorithm}\label{sec:algorithm}

Gaussian processes, typically denoted as $\mathcal{GP}(m(t), k(t, t'))$, with $t$ and $t'$ referring to two times of observation, are stochastic processes defining distributions over functions, such that for any finite collection of inputs the corresponding function values have a joint multivariate Gaussian distribution, i.e.
\begin{equation}
    \mathbf{y} \sim \mathcal{N}(\mathbf{m}, \mathbf{K}) 
    \label{eq:joint_prob}
\end{equation}
where $m_i=m(t_i,\boldsymbol{\theta})$ is the mean function and $\mathbf{K}_{ij}=k(t_i,t_j,\boldsymbol{\phi})$ where $\mathbf{K}_{ij}$ is the covariance matrix, while $k(t_i,t_j,\boldsymbol{\phi})$, thought of as a function over continuous $(t,t')$,  is the covariance function (or kernel) . Here, $\mathbf{t} = [t_1, \ldots, t_n]$ is the set observation timestamps, while $\boldsymbol{\theta, \ \phi}$ are the so-called process hyperparameters. The covariance function of a GP is defined so that the associated $\boldsymbol{K}$ is symmetric (i.e. $\mathbf{K}(t_i,t_j)=\mathbf{K}(t_j,t_i)$) and positive semi-definite.

GPs great advantage over alternative models is that, once a set of data points is collected, posterior predictions over the observed domain are readily available analytically:
\begin{equation}
\begin{split}    
\boldsymbol{\mu}_* &= \rm{\boldsymbol{K}}(\mathbf{t}_*, \mathbf{t}) \, [\boldsymbol{K}(\mathbf{t}, \mathbf{t}) + \sigma_n^2 \textbf{I}]^{-1} \, \mathbf{y} \\
\boldsymbol{\Sigma_*} &= \rm{\boldsymbol{K}}(\mathbf{t}_*, \mathbf{t}_*) - \boldsymbol{K}(\mathbf{t}_*, \mathbf{t}) \, [\boldsymbol{K}(\mathbf{t}, \mathbf{t}) + \sigma_n^2 \textbf{I}]^{-1} \, \boldsymbol{K}(\mathbf{t}, \mathbf{t}_*),
\end{split}
\label{eq:posterior_predictives}
\end{equation}
where $\boldsymbol{t_*}$ is the vector containing the times at which we want to predict the values of the light curve, $\boldsymbol{t}$ is the vector of the times used to fit the hyperparameters of the kernel, $\rm{\boldsymbol{K(\mathbf{t}, \mathbf{t})}}$ is the kernel computed on the points used to fit the GP, $\rm{\boldsymbol{K(\mathbf{t_*}, \mathbf{t})}}$ is the covariance between $\boldsymbol{t}$ and $\boldsymbol{t}_*$, $\sigma_n$ is the photometric error of each data point, $\boldsymbol{\mu_*}$ is the predicted mean, $\boldsymbol{\Sigma_*}$ is the prediction's covariance and finally $\boldsymbol{I}$ is the identity matrix. 

In this work, we use GPs only to fit the best hyperparameters describing the noise and the periodic signal, rather than for predictive purposes.

Once the mean and covariance functions are specified (e.g. for stationary Gaussian processes, the covariance function is simply related via the Fourier transform with the power spectral density, see \citealt{celerite}), one can fit the set of hyperparameters $\boldsymbol{\theta, \ \phi}$ that describe the data $D$. The log-likelihood of the data is given by \citep[see][]{Rasmussen2006}:
\begin{equation}
\begin{split}
\log p(D \mid M, \boldsymbol{\theta}, \boldsymbol{\phi}, I) &= 
-\frac{1}{2}\log |\boldsymbol{\Sigma}(\boldsymbol{\phi})| -\frac{N}{2}\log(2\pi) +
\\
&-\frac{1}{2} (D - \boldsymbol{m(\boldsymbol{\theta})})^{T} \boldsymbol{\Sigma}^{-1} (D - \boldsymbol{m}(\boldsymbol{\theta}))
\end{split}
\label{eq:GP_likelihood}
\end{equation}
where $N$ is the number of observations and $\boldsymbol{\Sigma}_{ij}=\boldsymbol{K}_{ij}+\sigma_i \ \boldsymbol{\delta_{ij}}$, where $\boldsymbol{K}$ and $\boldsymbol{m}$ are the covariance matrix and mean function, $\boldsymbol{\delta_{ij}}$ is the Kronecker delta and $\sigma_i$ denotes the observation noise variance at input $t_i$.

The evidence (or marginal likelihood)\footnote{From a statistical point of view, the log-likelihood in Eq.~\eqref{eq:GP_likelihood} is in fact the marginal one for a single GP model, while $Z$ is the (hyper-)evidence of each chosen GP family, e.g., the three specified in Secs.~\ref{sec:exp-kernel},~\ref{sec:cos-kernel}, and~\ref{sec:periodic_kernel}. We opt for a slight misnaming as it is ubiquitous in relevant literature.} of a particular model given a set of data is defined as:
\begin{equation}    
Z=\int {\rm d}\boldsymbol{\theta} p\left(D|M,\boldsymbol{\theta}\right) p \left(\boldsymbol{\theta}\mid M\right)  ,
\end{equation}
where $p(D|M,\boldsymbol{\theta})$ is the likelihood function introduced in Eq.~\eqref{eq:GP_likelihood} and $ p (\boldsymbol{\theta}|M)$ denotes the prior. 
In our context, the data $D$ consists of the observed light curve ($t_i$, $M_i$), with $i=1,2,...N$ labelling the datapoints and $M_i$ being the magnitude observed at time $t_i$. 
Finally, $M$ is the considered model, and $\boldsymbol{\theta}$ are the hyperparameters associated with it. 
Within the above formalism, the outcome of a model comparison is quantified by the so-called Bayes factor, i.e. the evidence-ratio between two competing models, $B=Z_{\rm{model,1}}/Z_{\rm{model,2}}$.
As we shall see in Section~\ref{subsec:Nested_sampling}, it strongly depends on the assumed kernel in each model.

Below, we briefly introduce three kernels, used to describe either random intrinsic AGN variability (`exponential kernel') or periodicities (`cosine' and `periodic' kernels, the latter being well-suited to periodic signals of arbitrary shape). 
To model both the noise and the periodic signal in the light curve, we construct kernels as the sum of periodic and exponential kernels.

\subsection{Exponential kernel}\label{sec:exp-kernel}
The intrinsic AGN variability in the optical and UV bands is well described by Damped Random Walk \citep[DRW, see][]{Kelly2009, MacLeod10, Kozlowski2010} and is parameterised by a damping timescale $\tau$, which represents the light curve autocorrelation timescale, and a variability amplitude, $\hat{\sigma}$, which quantifies the typical magnitude of intrinsic variations. Therefore, the process power spectral density (PSD) reads:
\begin{equation}
    \mathrm{PSD}(f)= \frac{2 \hat{\sigma}^2 \tau^2}{1+(2 \pi \tau f)^2} \ ,
    \label{eq:DRW_power}
\end{equation}
where $f$ denotes the frequency. Such a stochastic process can be equivalently described by the kernel: 
\begin{equation}
    k_{\rm{DRW}}(t_i,t_j)=\frac{1}{2}\hat{\sigma}^2 \tau \exp \left(-\frac{|t_i-t_j|}{\tau}\right)
    \label{eq:DRW_covariance} .
\end{equation}

\subsection{Cosine kernel}\label{sec:cos-kernel}

GPs can be used to model periodic signals. This can be done by employing a periodic mean function (e.g. $m(t)=A\sin(2\pi t/P)$) ---with $P$ being its fundamental period--- and modelling the noise through the covariance matrix, e.g. using the kernel defined in Equation~(\ref{eq:DRW_covariance}) to describe red noise. Alternatively, the periodic component can be directly incorporated in the noise term of the covariance matrix. 
In this work, we will follow the second approach, as a deterministic analytical description of the expected modulation from the interaction with the gas surrounding the binary is still unavailable, and assume a constant mean function equal to 0 ($\boldsymbol{m(\theta)}=0$).

Sinusoidal signals can be modelled using the cosine kernel (see Section~\ref{sec:recoveries}), which reads:
\begin{equation}
    k_{\rm{cosine}}(t_i,t_j)=A^2 \cos\left(2\pi\frac{|t_j-t_i|}{P}\right)
    \label{eq:cos_kernel}
\end{equation}
where $A$ is the autocorrelation strength, and $P$ is its period. 

As the correlation between two photometric points with $\Delta t \ll P$ is always close to 1 by construction, the cosine kernel fails in properly modelling sudden jumps in the light curves on timescales shorter than $P$. 

\subsection{Generic periodic kernel}\label{sec:periodic_kernel}
In this work, we employ a more generic periodic kernel, flexible enough to model periodic signals of arbitrary shapes, defined by:
\begin{equation}
    k_{\rm{periodic}}(t_i, t_j) = A^2 \exp\left(-\frac{2\sin^2\left(\frac{\pi}{P} | t_i - t_j | \right)}{l^2}\right)
    \label{eq:periodic_kernel}
\end{equation}
The hyperparameter $l$ determines how rapidly correlations decay as the distance between two points increases within a period.
We note that a similar kernel was used to study stellar variability, exoplanetary science and eclipsing binary characterisation \citep[see][and references therein for applications]{Haywood2014, Vanderburg2015, Angus2018, Gonzalez2021, Wang2025}. However, in such works, the kernel is quasi-periodic (the periodic kernel is multiplied by an exponential decay) rather than the purely periodic kernel we are assuming here.

Unlike the cosine kernel, whose Fourier expansion over $\Delta t$ consists of a single Fourier component at frequency $f=1/P$, limiting it to a single sinusoidal term, the periodic kernel can be expanded in an infinite Fourier series over $\Delta t = \left|t_i-t_j\right|$. For this reason, it can describe periodicities with arbitrary shapes.

The contribution of each harmonic is weighted by the kernel lengthscale $l$: smaller values of $l$ upweight higher harmonics and allow for sharp edges in the signal, while larger values of $l$ downweight them, yielding smoother, nearly sinusoidal signals. 
A more detailed discussion on this point is reported in Appendix~\ref{app:spec_periodic}. 
By having a higher-dimensional parameter space and a different functional form, the generic periodic kernel is suitable to describe signals with a larger complexity, as compared to the cosine one. 
Therefore, it is only through Bayesian model selection that we will be able to assess whether such complexity can find support in observed data.

\begin{figure}
    \centering
    \includegraphics[width=\linewidth]{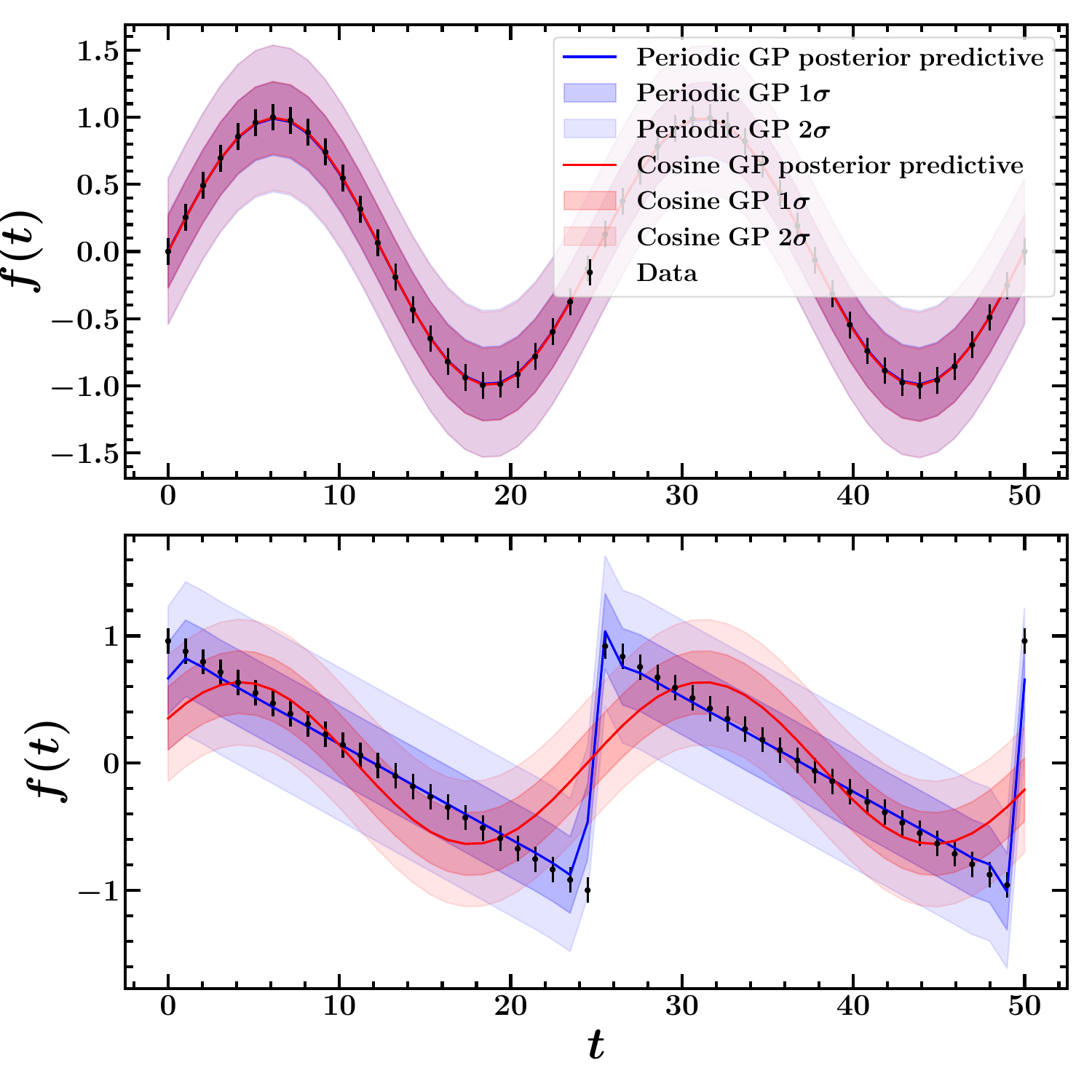}
    \caption{Posterior predictive distributions from GP inference on a sinusoidal light curve (upper panel) and a sawtooth light curve (lower panel). 
    Results from the cosine kernel (Equation \ref{eq:cos_kernel}) are shown in red, and from the periodic kernel (Equation \ref{eq:periodic_kernel}) in blue. 
    Shaded regions indicate the $1\sigma$ (dark) and $2\sigma$ (light) credible intervals. 
    The solid lines and shaded regions in the upper panel clearly show that both the periodic and cosine kernels can reproduce the sinusoidal signal as they overlap. The bottom panel, instead, highlights both the inadequacy of the cosine kernel in modelling non-sinusoidal light curves and the flexibility of the periodic kernel that allows it to describe periodicities with arbitrary shapes.}
    \label{fig:fit_examples}
\end{figure}

Figure \ref{fig:fit_examples} shows the posterior predictive distributions, defined in Equation (\ref{eq:posterior_predictives}), for idealised, mock signals: a sinusoid (upper panel) and a sawtooth signal (lower panel), evenly sampled with a daily cadence. 
The two examples are fitted using both the cosine kernel (Eq.~\ref{eq:cos_kernel}) and the periodic kernel (Eq. ~\ref{eq:periodic_kernel}). Both kernels accurately describe the sinusoidal light curve. For the sawtooth light curve, both kernels recover the correct period, but only the periodic kernel captures the true shape of the signal. This distinction is important for model comparison, as the Bayes factor in favour of the cosine kernel in the sawtooth case would be suppressed. 
Specifically, we find $\log_{10} (Z_{\mathrm{periodic}}/Z_{\rm{cosine}}) \sim -5$ for the sinusoidal light curve, as expected, since both models accurately describe the sinusoidal shape, but the cosine kernel requires a parameter less than the periodic kernel. On the other hand, for the sawtooth light curve, we get $\log_{10} (Z_{\mathrm{periodic}}/Z_{\rm{cosine}}) \sim 100$, highlighting the fact that the cosine kernel is not suitable to describe such signals. In the following analysis, we adopt the generic periodic kernel, while comparisons with the cosine kernel and with the LSP analysis are discussed in Sections~\ref{sec:recoveries} and~\ref{sec:LSP_comparison}, respectively.

\subsection{Model selection} \label{subsec:Nested_sampling}
In this work, we search for the GP hyperparameters through Bayesian inference, obtaining posterior samples with nested sampling \citep[see][]{Skilling2006}. Nested sampling is a statistically robust algorithm suitable for the estimation of the model parameters (or hyperparameters in the context of GPs) and the associated uncertainty, in the form of posterior samples. In addition, it directly provides an estimate of the model evidence. More specifically, we use \texttt{Gpytorch} \citep{gardner2021} to construct the GP kernel and to evaluate the likelihood. For the nested sampling algorithm, we use the \texttt{raynest} implementation \citep[see][]{CPnest} with 500 live points, and the stopping is set so that the evaluation stops as soon as the increment of the log-evidence is smaller than $\ln Z< 0.1$.

In the following, we compare two models: a pure noise model given by the exponential kernel defined by Equation (\ref{eq:DRW_covariance}) and a model that includes both a periodic modulation and noise, described by the sum of kernels in Equations (\ref{eq:DRW_covariance}) and (\ref{eq:periodic_kernel}).

To ensure that the priors are sufficiently broad for proper posterior sampling across all objects, we rescale the light curves. Specifically, we subtract the mean and divide by the standard deviation of each quantity. This is done for both times and magnitudes of the light curve data. \footnote{We note that rescaling the time axis by the standard deviation of the observation times is effectively similar to rescaling by the total duration of the light curve. Such normalisation prevents us from exploring timescales that are much longer than the observed baseline.}

In this work, we assume log-uniform priors for all the rescaled GP hyperparameters.

\vspace{0.4em}
\noindent \textbf{Noise model:}
\vspace{-1.0em}
\begin{align*}
    \tau &\sim \text{LogUniform}(10^{-4},\ 10^{2}) \\
    \hat{\sigma} &\sim \text{LogUniform}(10^{-4},\ 10^{2})
\end{align*}

\vspace{1em}

\noindent \textbf{Periodic model:}
\begin{align*}
    \tau &\sim \text{LogUniform}(10^{-4},\ 10^{2}) \\
    \hat{\sigma}&\sim \text{LogUniform}(10^{-4},\ 10^{2}) \\
    P &\sim \text{LogUniform}(10^{-2},\ 10^{2}) \\
    l &\sim \text{LogUniform}(10^{-6},\ 10^{6}) \\
    A &\sim \text{LogUniform}(10^{-3},\ 10^{2})
\end{align*}

We calculate the Bayes factor between the periodic+noise and noise-only models to select the best model describing the data. 

In this work, we fix the Bayes factor threshold to claim the detection of a periodicity by determining the distribution of the Bayes factor produced by noise-only realisations (see Section~\ref{sec:threshold}).

\subsection{Simulations of mock light curves}\label{sec:lightcurves}
In this work, we analyse light curves consisting of a periodic signal superimposed on realistic quasar noise (modelled as a DRW). The periodic signal is characterised by three parameters: the period $P$, the amplitude $A$, and the phase $\phi$, while the noise is described by a damping timescale $\tau$ and an amplitude $\sigma$, for a total of 5 parameters used to generate each light curve.

We use the same light curves used in \cite{Lin2025}, where periodicity-related parameters were uniformly sampled from [0, 0.5), [0,$T_{\rm data}/1.5$), and [0, 2$\pi$) for $A$, $P$, and $\phi$, respectively, and with $T_{\rm{data}}$ being the baseline of the respective PTF lightcurve. 
Parameters associated with the intrinsic AGN variability are derived from each quasar properties based on the correlations from \citealt{MacLeod10} \citep[see][for full details]{Lin2025}.

For each set of input parameters, 4 types of light curves are generated. The periodic component can have either a sinusoidal or a sawtooth-like shape, while the observational baseline is constructed with different cadences: PTF-like (with the same cadence as the one of the Palomar Transient Factory) or idealised (sampled daily).
The PTF-like light curves are constructed using the same cadence as those analysed in \cite{Charisi2016}. We assign to each point photometric errors sampled from a Gaussian distribution with zero mean and a standard deviation equal to the photometric uncertainty of the corresponding point in the real PTF light curve.  
For the idealised light curves, photometric errors are added analogously, though with standard deviation equal to the mean photometric error of the corresponding PTF light curve \citep[see][for additional details]{Lin2025}. 
The typical mean photometric error associated with real PTF light curves is $\sim 0.05$.

We complement our analysis with 2000 additional LSST-like light curves from \citet{Lin2025} (1000 with sinusoidal variability and 1000 with sawtooth luminosity profiles). We select this sub-set of light curves based on the analysis of the PTF-like light curves, for which the periodicity was not detected.

\begin{figure}
    \centering
    \includegraphics[width=\linewidth]{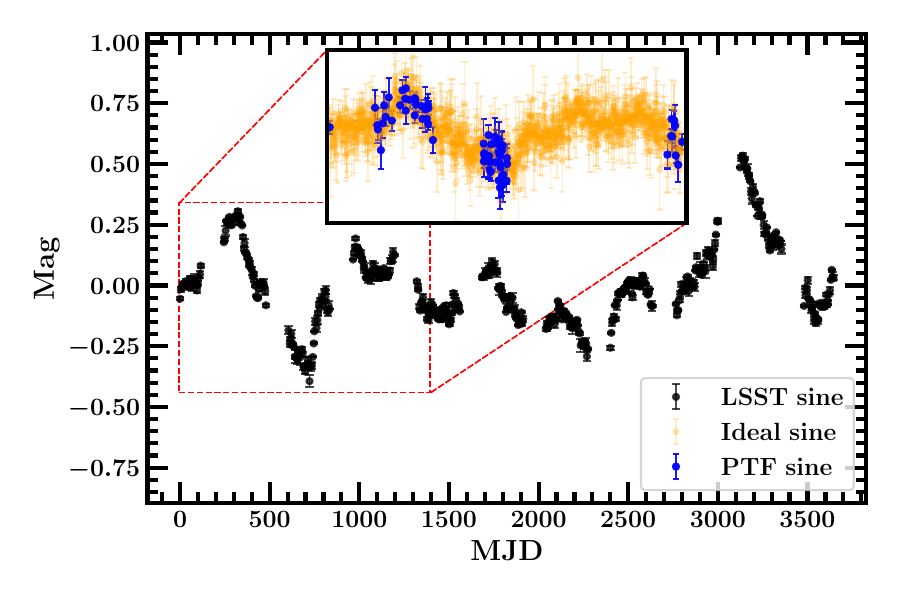}
    \caption{Examples of the same sampled sinusoidal light curve for the three baselines. The LSST light curve is shown in the main panel, while the PTF (blue points) and ideal (orange points) light curves are displayed in the inset (lower right). The dashed red rectangle in the main panel highlights the typical duration of the PTF and ideal sampling, emphasising the longer observational coverage of the LSST baseline.}
    \label{fig:example_lightcurves}
\end{figure}

These light curves were generated using the same parameters as in the PTF-like ones but assuming LSST-like cadence, magnitude uncertainties, and an extended baseline of equal to the expected survey duration, $T_{\rm{obs}}\sim 10 \ \rm{yr}$. Specifically, LSST-like light curves have a median cadence of around $5$ days, with seasonal gaps of approximately 4 months each year, and show a median uncertainty of $6.5 \times 10^{-3}\ \rm{mag}$, an order of magnitude smaller compared to the PTF-like and ideal ones. 
In Figure~\ref{fig:example_lightcurves}, we show one example light curve for each baseline (the LSST baseline in the main plot and the ideal and PTF baselines in the inset).
We emphasise that we limit ourselves to only 1000 LSST-like light curves for each periodic signal (instead of the total 12,400 analysed in \citealt{Lin2025} and above) due to the computational cost of the analysis; a complete analysis of a single LSST-like light curve takes up to 8 hours. In Section~\ref{sec:conclusions}, we discuss possible ways to speed up the code, in order to address this significant limitation of our method, which does not allow us to scale our calculations to large quasar samples.

\section{Results}\label{sec:results}

In this section, we present the results of the analysis of 12400 light curves for each combination of short baselines (PTF-like and idealised) and shape of the periodicity (sinusoidal and sawtooth), as well as a subsample of 1000 LSST-like light curves for each periodicity shape.

\subsection{Bayes factor threshold for periodicity detection}\label{sec:threshold}
To assess whether a light curve shows significant evidence for periodicity, it is necessary to set a threshold in the Bayes factor obtained by the comparison between the noise-only and the periodic+noise models. In this work, we set this threshold in two ways. The first is based on fixing the false positive rate. Specifically, we set this threshold by finding a Bayes factor threshold based on the false positive rate assumed in \cite{Lin2025} to make a fair comparison discussed in Section~\ref{sec:LSP_comparison}. The second is based on the ROC curve. Specifically, we set this threshold by finding the value of the Bayes factor that maximises the distance from the curve describing random classification. The main results discussed in this work are obtained through the ROC-based threshold, as it maximises the efficiency of periodicity detection by keeping the false positive rate (the fraction of non-periodic signals being misidentified as periodic) low.
We characterise the FAP, generating $4 \times 12400$ noise-only light curves ($12400$ light curves for each of the four combinations of sinusoidal and sawtooth signals for the idealised and PTF-like data). Specifically, we compute the Bayes factor between the noise-only and periodic+noise models for all the noise-only light curves and get their distribution.

\begin{figure}[t!]
    \centering
    \includegraphics[width=\linewidth]{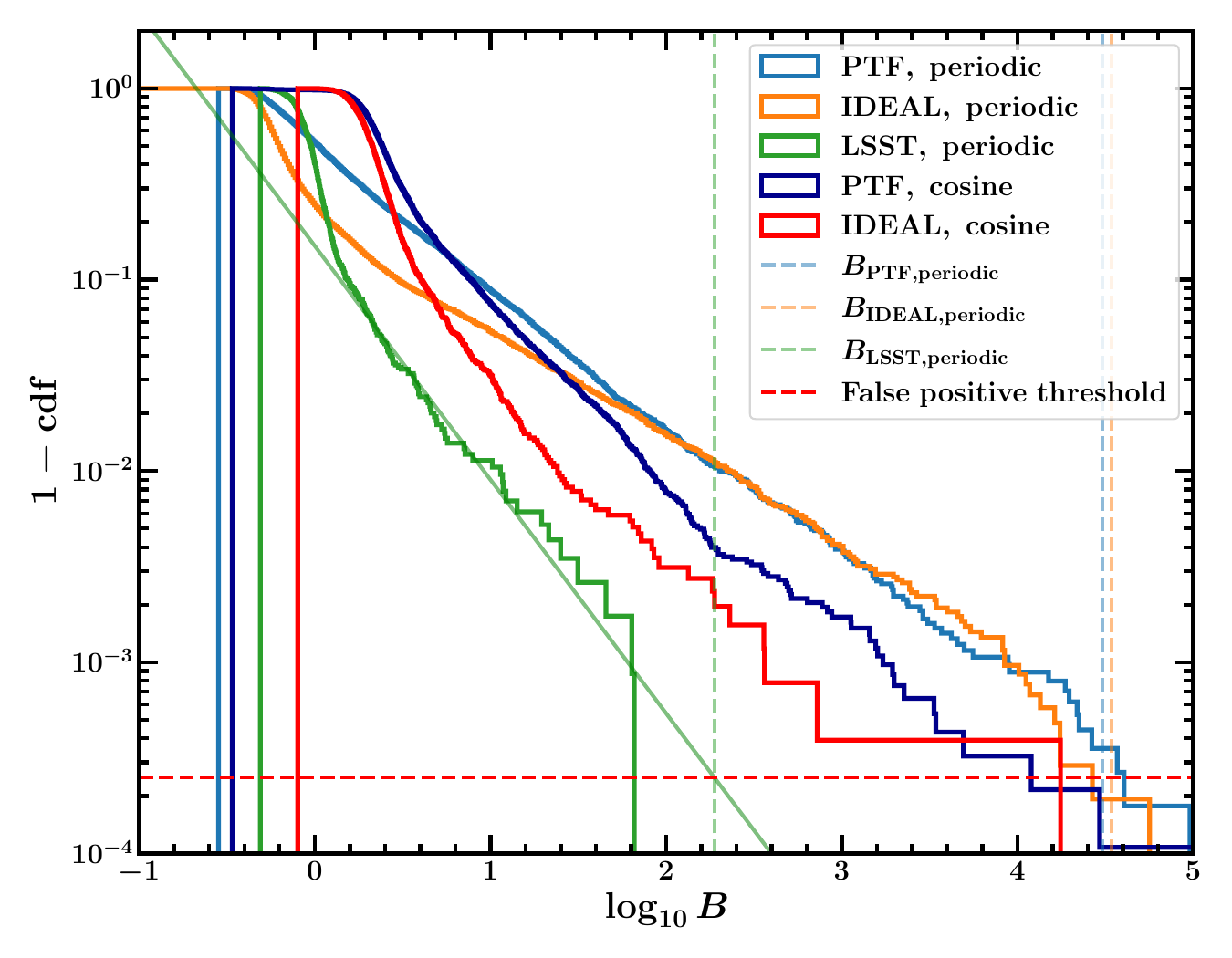}
    \caption{Distribution of the Bayes factors for the $10^4$ noise-only light curves for the PTF using the periodic kernel and cosine kernel (blue and dark blue distributions) and ideal using the periodic kernel and cosine kernel (orange and red distributions) baselines, and the $10^3$ noise-only light curves for the LSST (green distribution) baseline found by analysing the light curves with the generic periodic kernel. The vertical lines refer to the Bayes factor threshold identified for a fair comparison with the LSP analysis discussed in more detail in Section~\ref{sec:LSP_comparison}.} 
    \label{fig:threshold}
\end{figure}

In Figure~\ref{fig:threshold}, we show the cumulative distribution of $\log_{10}B$ for all three sampling scenarios for the periodic kernel: PTF-like (blue distribution), idealised (orange distribution), and LSST-like (green distribution). 

From the distributions in Figure~\ref{fig:threshold} we retrieve the threshold on the Bayes factor to obtain a FAP<$25\times 10^{-5}$ that will allow with the LSP analysis (see Section~\ref{sec:LSP_comparison}) \footnote{The multiplicative factor 25 comes from the fact that when analysing the LSP, \cite{Lin2025} makes use of 25 frequency bins}. Such thresholds are found between $\log_{10}B=4-5$ for the ideal and PTF-like light curves and between $\log_{10}B=2-3$ for LSST-like light curves. To be conservative, we assume a Bayes factor threshold of $\log_{10}B=5$ for ideal and PTF-like light curves and of $\log_{10}B=3$ for LSST-like light curves when comparing our results with the LSP-analysis presented in Section~\ref{sec:LSP_comparison}.

\begin{figure}[t!]
    \centering
    \includegraphics[width= \hsize]{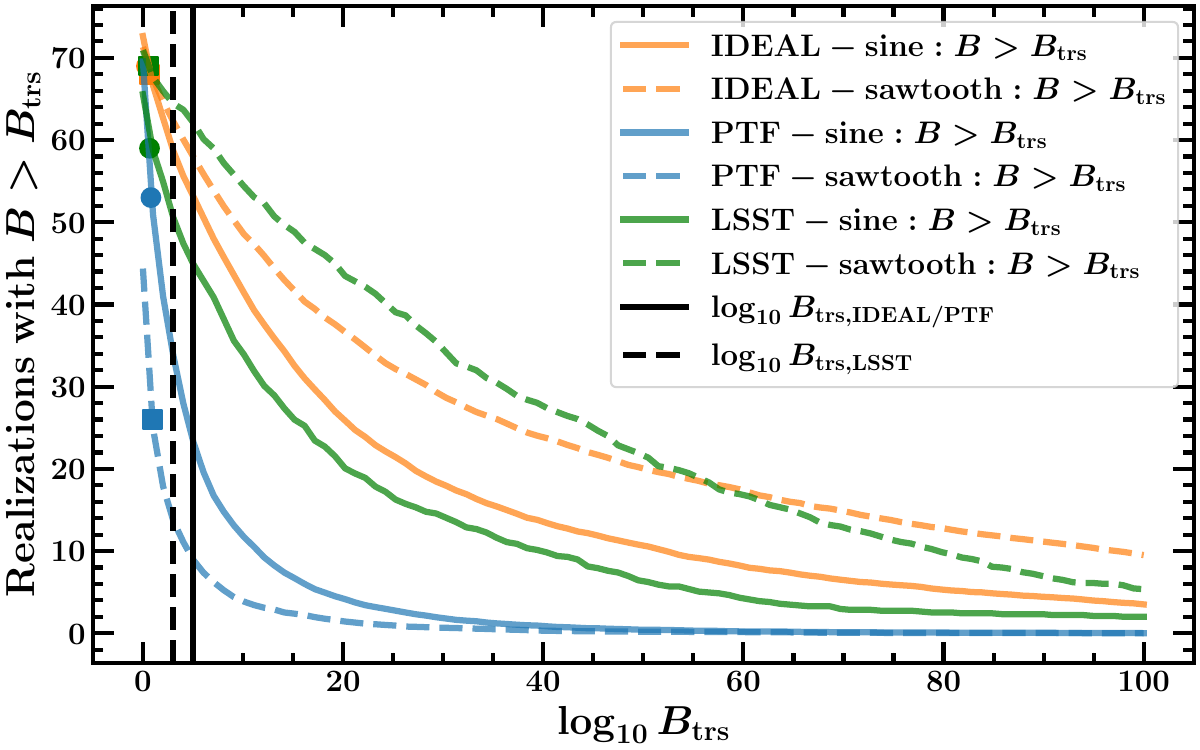}
    \caption{Fraction of realisation containing periodic signals with a Bayes factor greater than a threshold as a function of the assumed threshold. Solid lines refer to the sinusoidal light curves, while dotted lines refer to sawtooth light curves. Blue, red and green colours refer to the ideal, PTF and LSST baselines, respectively. The vertical lines refer to the Bayes factor threshold of $\log_{10} B_{\rm{trs}}=5$ for the ideal, and PTF baselines (solid line) and $\log_{10} B_{\rm{trs}}=3$ for the LSST baseline (dashed line) used to compare the results with the LSP-analysis, see Section~\ref{sec:LSP_comparison}. Circles and squares identify the points at the identified ROC curve-based Bayes threshold for the sinusoidal and sawtooth cases, respectively. Orange markers refer to Ideal-like light curves, blue markers refer to PTF-like light curves and finally, green markers refer to LSST-like light curves.}
    \label{fig:percentages_threshold.pdf}
\end{figure}

The true-positive fraction is computed by dividing the number of periodic light curves correctly identified as periodic by the total number of periodic light curves analysed. These fractions are shown in Figure~\ref{fig:percentages_threshold.pdf}.

\begin{figure}[t]
    \centering
    \includegraphics[width=\linewidth]{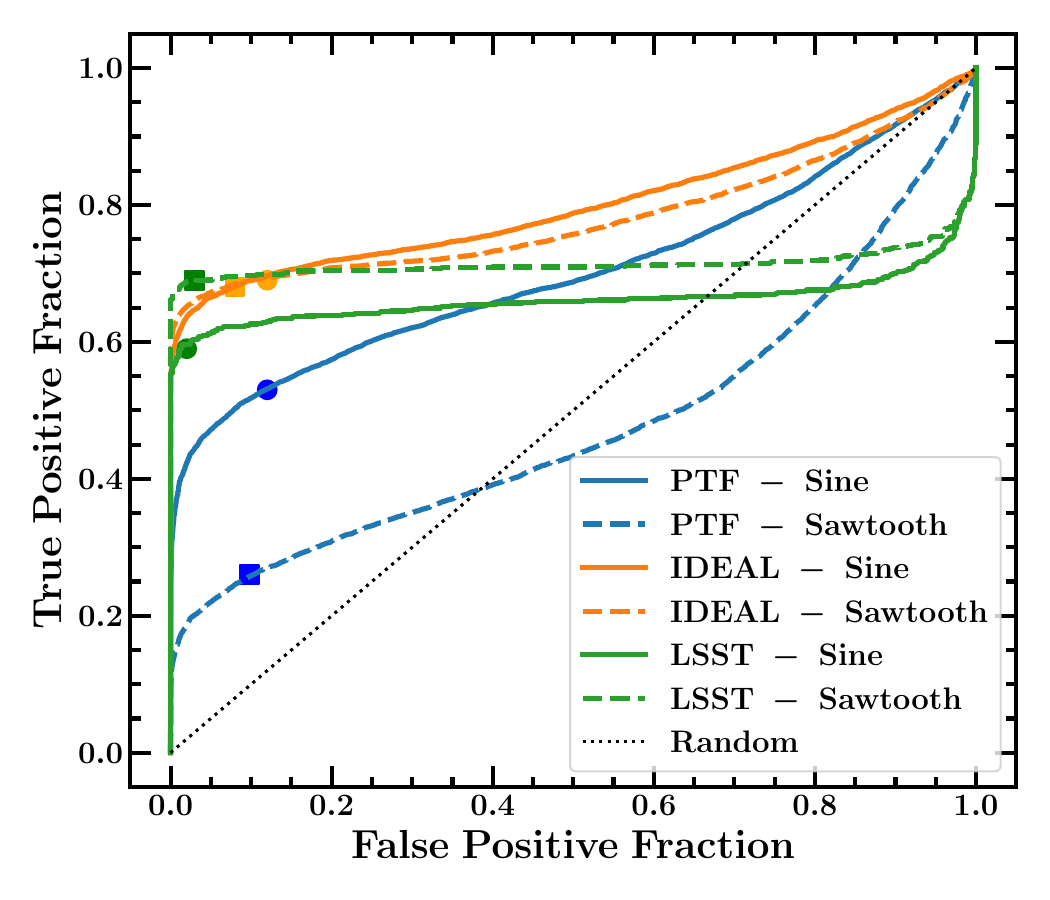}
    \caption{ROC curves of the different combinations of baselines and periodicity shape. Solid lines refer to sinusoidal light curves, while dashed lines refer to sawtooth light curves. Blue lines refer to the PTF-like baseline, orange lines refer to the idealised baseline, while green lines refer to the LSST baseline. Circles and squares identify the points at the identified Bayes threshold for the sinusoidal and sawtooth cases, respectively. Blue markers refer to PTF-like light curves, orange markers refer to ideal light curves and finally, green markers refer to LSST-like light curves. }
    \label{fig:ROC_curves}
\end{figure}

The distributions in Figures~\ref{fig:threshold} and~\ref{fig:percentages_threshold.pdf} allow for the construction of the Receiver Operating Characteristics curve \citep[ROC, see][]{Marcum1947}, which we use to identify an optimal value of the threshold, by finding the point where the value of $d(TP)/d(FP)=1$ (i.e. the farthest point from the $TP=FP$ curve which indicates a random guess \footnote{This can be demonstrated by considering a point ($FP_0$, $TP_0$) and the one-to-one line $TP=FP$. The distance between the point and the line is given by $(TP-FP)/\sqrt{2}$. Maximising this distance implies $d(TP-FP)/dFP =0$, thus  $dTP/dFP =1$}) with $TP$ and $FP$ being the true positive and false positive rates, respectively. The ROC curves are shown in Figure~\ref{fig:ROC_curves} and the thresholds for $\log_{10}B$ in the ideal-sine, ideal-sawtooth, PTF-sine, PTF sawtooth, LSST-sine and LSST-sawtooth are $0.35$, $0.64$, $0.80$, and $0.94$, $0.66$ and $0.56$, respectively. We note that the different thresholds between the baselines are mainly due to the quality of the data as well as the different ability of the red noise to reproduce in the same way both shapes (sinusoidal and saw-tooth) of the periodic signal. Interpreting these thresholds by means of the Jeffreys scale \citep[see][]{jeffreys1998}, we would get substantial evidence in favour of the periodic model.

\subsection{Recovery fractions and parameter recovery} \label{sec:recoveries}

The true positive fraction and false positive fractions obtained using the ROC-based thresholds for each baseline and variability shape are summarised in Table~\ref {tab:percentages_ROC}.

\begin{table}[h!]

    \centering
    \caption{Fractions of retrieved periodicities using the thresholds from the ROC curves}
    \begin{tabular}{|c|c|c|c|c|}
        \hline
        & \multicolumn{2}{|c|}{ True Positive}& \multicolumn{2}{|c|}{False Positive }\\
        \hline
         Periodic kernel& Sine & Saw  & Sine  & Saw \\
         \hline
        idealised & 69.7$\%$ & 68.4$\%$ &12.0$\%$ & 8.1 $\%$\\
         \hline
        PTF-like & 53.4$\%$ & 25.7$\%$ & 12.5 $\%$& 9.8$\%$\\
         \hline
        LSST-like & 59.3$\%$ & 69.0$\%$ & 2.1$\%$& 3.1$\%$\\
         \hline
        Cosine kernel& Sine & Saw  & Sine  & Saw \\
        \hline
        idealised & 55.2$\%$ & 8.5$\%$ &1.3$\%$ & 1.5 $\%$\\
         \hline
        PTF-like & 17.1$\%$ & 2.9$\%$ & 6.5 $\%$& 10.4$\%$\\
         \hline
    \end{tabular}
    \tablefoot{Fraction of light curves yielding a Bayes factor larger than the thresholds found through the ROC curves for all the combinations of baselines and shapes of the periodic signal using both the generic periodic and cosine kernels.}
    \label{tab:percentages_ROC}
\end{table}

From our analysis, we find that the true positive fractions are similar for the sinusoidal and sawtooth scenarios when considering the idealised baseline being  $\sim68\%$. We observe that the efficiency of our algorithm decreases as the number of observational data points drops. Specifically, only$\approx 53\%$ ($\approx 26\%$) of PTF-like light curves are identified as periodic in the sinusoidal (sawtooth) case. We interpret the stronger impact of fewer and sparser data on the sawtooth light curves as a consequence of their strong dependence on whether the fast-rising phase (the key signature distinguishing them from red noise) is included in the observations or lost in the gaps. \footnote{Such an occurrence is rarer the longer the light curve is, as it is unlikely that all peaks would fall in gaps. The analysis of LSST-like data presented below further supports this interpretation.}

Finally, for the LSST-like light curves, we see that the retrieved fractions are closer to the ones of the idealised case, being slightly reduced at $59\%$ for the sinusoidal scenario, regardless of the presence of quasi-periodic observational gaps and of the LSST-like light curves being a subset of PTF-like ones, for which the periodic model was not favoured with respect to the noise-only model. 
As expected, we observe an overall trend: by increasing the number of datapoints or the number of observed cycles, the fraction of retrieved periodicities increases. We defer the discussion on the most relevant parameters for a correct identification of a periodic component to section~\ref{sec:param_importance}.

While our true positive fraction is much larger than those quoted by alternative searches in literature \citep[e.g.][]{Lin2025}, we stress that they were not obtained by fixing a Bayes factor threshold, as we do in our ROC curve analysis. In Section~\ref{sec:LSP_comparison}, we find a Bayes factor threshold to perform a fair comparison with the work by \cite{Lin2025} and show how the fraction of retrieved periodicities changes by using that threshold.

As in \cite{Lin2025}, we assess the goodness of the parameter recovery using the relative fractional error on the period as a metric. In particular, we define the metric of our test as:
\begin{equation}
    \delta P= \frac{P_{\rm{rec}}-P_{\rm{inj}}}{P_{\rm{inj}}} \times 100\%
    \label{eq:metric}
\end{equation}
\begin{table}[h!]
    \centering
    \caption{Fraction of retrieved periodicities with $|\delta P|<0.15$ using the thresholds from the ROC curves}
    \begin{tabular}{|c|c|c|}
        \hline
         & Sinusoid & Sawtooth \\
         \hline
        Idealised & 82.3 $\%$ & 99.8$\%$ \\
         \hline
        PTF-like & 61.1$\%$ & 69.4$\%$ \\
         \hline
        LSST-like & 92.0 $\%$ & 98.8$\%$ \\
         \hline
    \end{tabular}
    \tablefoot{Fraction of light curve with a retrieved period within $15\%$ of the injected one among the light curves with a Bayes factor greater than the detection thresholds found through the ROC curves for all the combinations of baselines and shape of the periodic signal.}
    \label{tab:fractions_recovered}
\end{table}

\begin{figure}[h!]
    \centering
    \includegraphics[width=\linewidth]{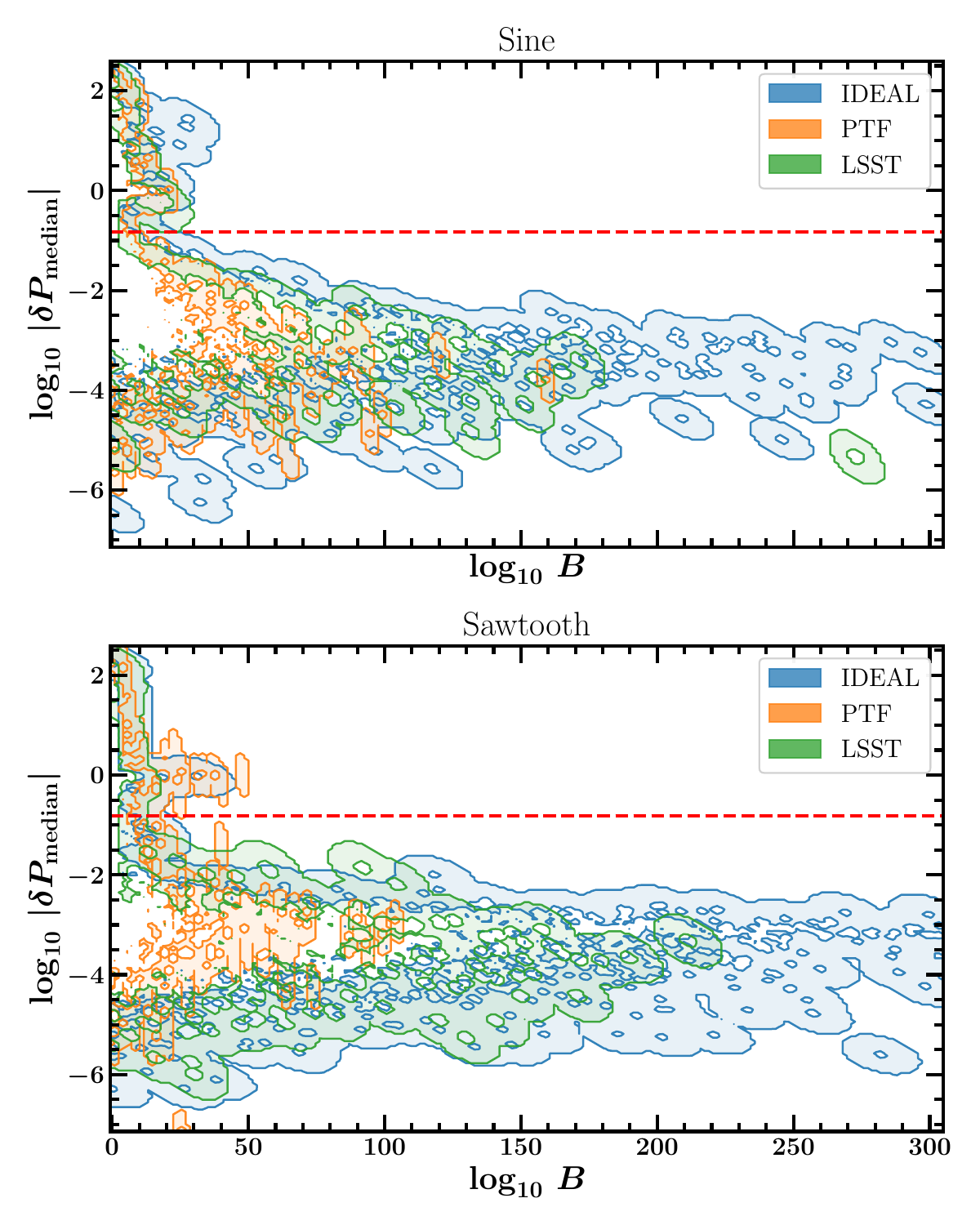}
    \caption{Absolute value of the relative period recovery error as a function of the base 10 logarithm Bayes factor, shown as density contours at the 25th and 75th percentile levels obtained through Gaussian kernel density estimation. Blue, orange and green markers refer to the ideal, PTF and LSST baselines, respectively. The horizontal-dashed red line refers to a reference value $|\delta P|=0.15$. The upper panel refers to sinusoidal light curves, while the lower panel refers to sawtooth light curves.}
    \label{fig:delta_P_vs_B_all}
\end{figure}

In Figure~\ref{fig:delta_P_vs_B_all}, we show the absolute value of the relative per cent error ($|\delta P|$) versus the Bayes factor as a scatter plot. In particular, it can be seen that the relative error tends to decrease as the Bayes factor increases.

We find that the periods are well retrieved in the idealised and LSST cases when we consider light curves with Bayes factors greater than the thresholds found through the ROC curves shown in Figure~\ref{fig:ROC_curves}. The test has more difficulties in recovering the correct period in the PTF light curves with respect to the other baselines. This is mostly due to the sampling sparseness. 
A particularly interesting case is that of light curves exhibiting a sawtooth periodicity. As shown by the orange diamonds in Figure~\ref{fig:delta_P_vs_B_all}, a horizontal feature appears at $|\delta P| = 1$. 
This occurs because, in these light curves, the sharp rise of the periodic signal is not observed in every cycle. The absence of this part of the signal leads the algorithm to identify a period that is twice the injected one. 

\subsection{Comparison with the cosine kernel}
Similar periodicity searches using nested samplers to explore the parameter space of Gaussian processes have been performed in the literature \citep[see][for example]{foustoul25}, but, as mentioned in Section~\ref{sec:introduction}, the cosine kernel as defined in Equation~(\ref{eq:cos_kernel}) was employed. 
Having a different number of free parameters, the distribution of the false positive fraction as a function of the assumed threshold on the Bayes factor is expected to be different. 
To study it, we analyse using the cosine kernel the same light curves used to build the distribution for the periodic kernel. The false positive distributions for the cosine kernel are shown in red (idealised light curves) and dark blue (PTF-like light curves) in Figure~\ref{fig:threshold}. 

\begin{figure}[t!]
    \centering
    \includegraphics[width=\linewidth]{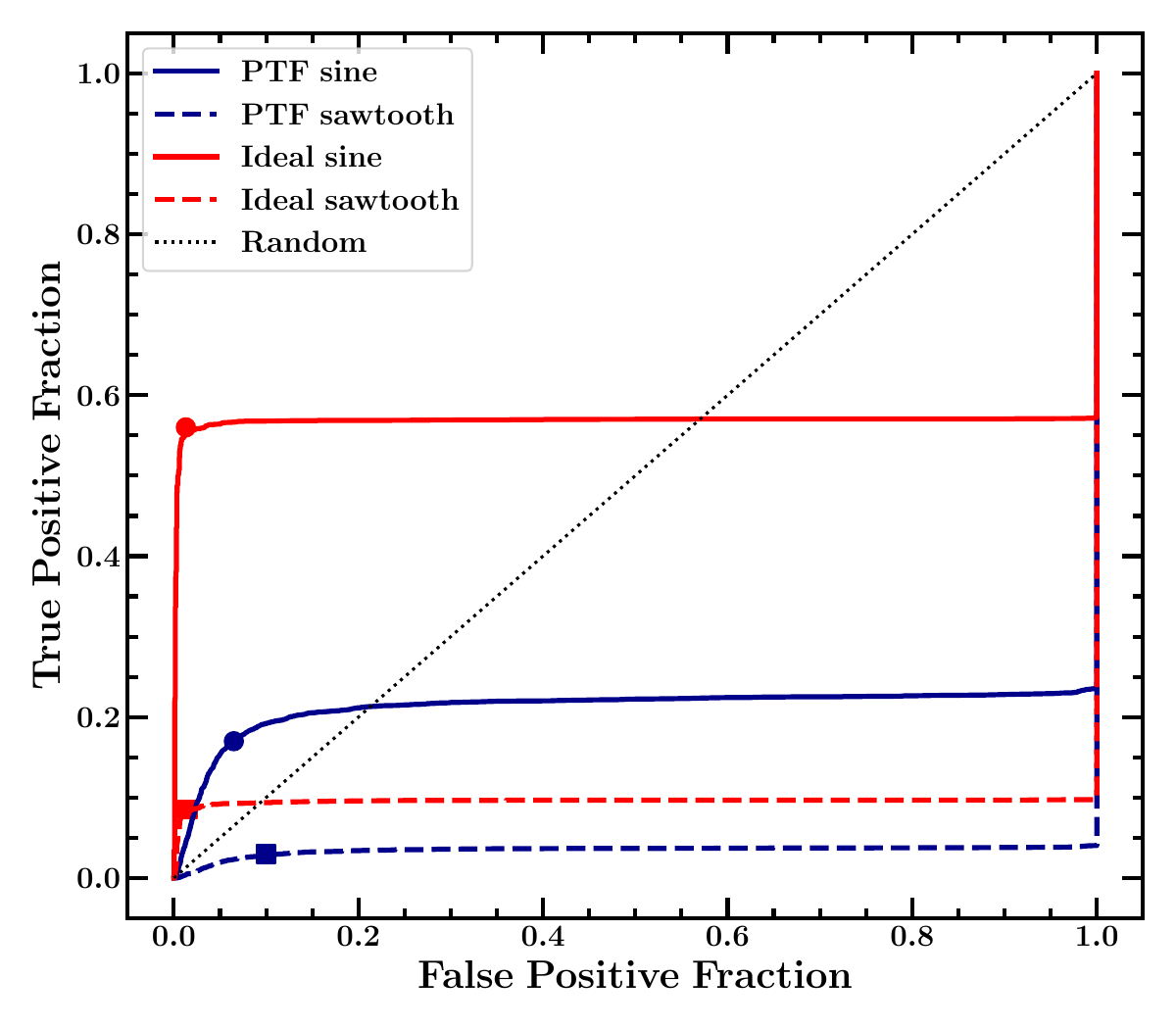}
    \caption{ROC curves of the different combinations of baselines and periodicity shape found using the cosine kernel. Solid lines refer to sinusoidal light curves, while dashed lines refer to sawtooth light curves. Blue lines refer to the PTF-like baseline, while red lines refer to the idealised baseline. Circles and squares identify the points at the identified Bayes threshold for the sinusoidal and sawtooth cases, respectively. Blue markers refer to PTF-like light curves, while red markers refer to ideal light curves. }
    \label{fig:thresholds_cosine}
\end{figure}
For the cosine kernel, the $\log_{10} B$ thresholds found through the ROC curves, shown in Figure~\ref{fig:thresholds_cosine}, are 1.29, 1.22, 1.06, 0.86 for the ideal sinusoidal, ideal sawtooth, PTF ideal and PTF sawtooth, respectively. 
The percentages of true and false positives found with the cosine kernel using these thresholds are summarised in the lower half of Table~\ref{tab:percentages_ROC}.

\begin{figure}[t!]
    \centering
    \includegraphics[width=\linewidth]{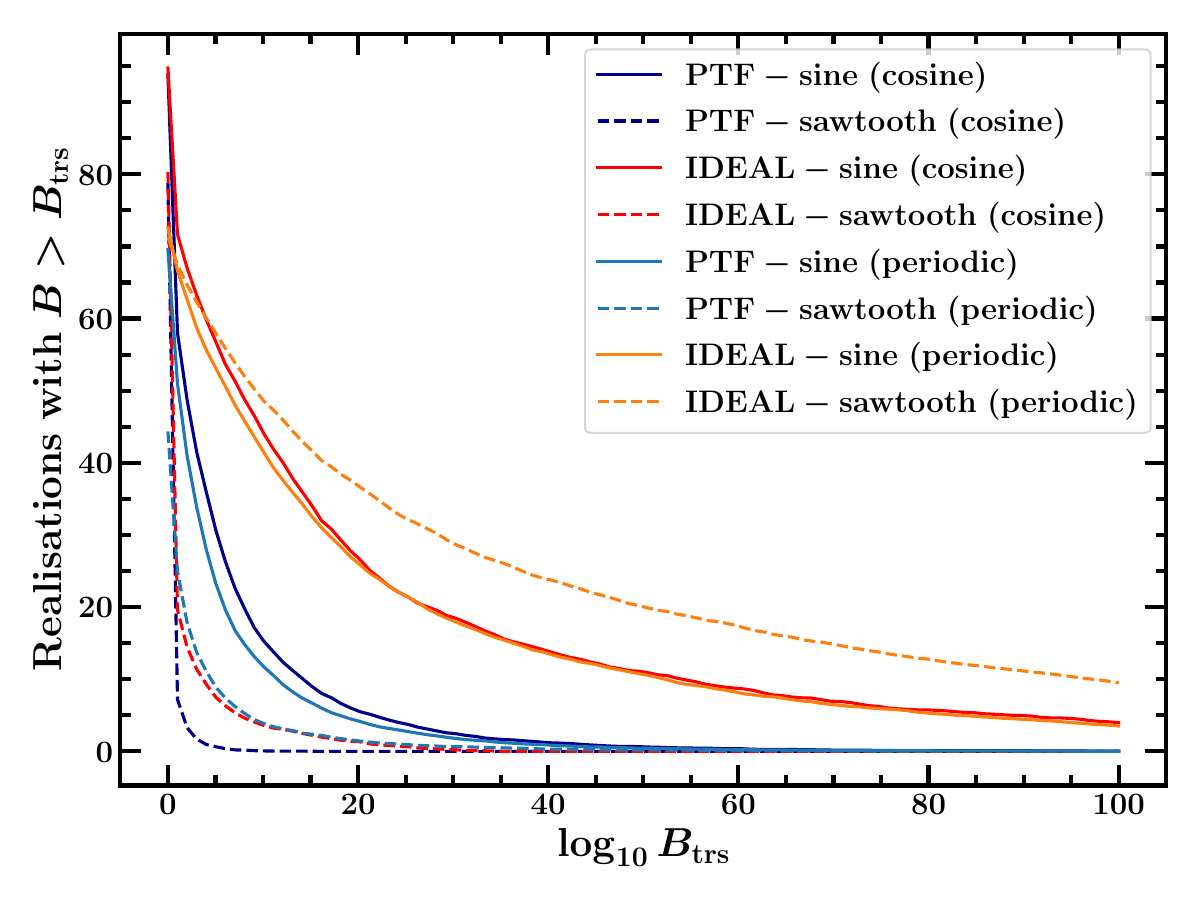}
    \caption{Comparison of the fraction of realisations with a Bayes factor greater than a threshold as a function of the assumed threshold found using the cosine or generic periodic kernels. Solid lines refer to the results obtained with the cosine kernel, while dotted lines refer to the results obtained with the generic periodic kernel. The colour of the lines identifies a different combination of baseline and periodicity shape: blue refers to PTF-like sinusoidal light curves, red refers to PTF-like sawtooth light curves, pink refers to Ideal sinusoidal light curves, and finally cyan refers to Ideal and sawtooth light curves.}
    \label{fig:percentages_cosine}
\end{figure}

The cosine kernel fails in finding periodicities that are not sinusoidal. In particular, for non-sinusoidal signals with sparse sampling, the fraction of false positives at the identified threshold is much higher than the true positive fraction. This is confirmed by Figure~\ref{fig:percentages_cosine}, where the ability of the cosine kernel to identify non-sinusoidal signals rapidly drops when compared to the generic periodic kernel.

From Table~\ref{tab:percentages_ROC}, the cosine kernel retrieves a lower fraction of true positives also in the idealised and sinusoidal scenario with respect to the fractions retrieved by the generic periodic kernel. This is mainly due to the different Bayes factor thresholds caused by the different shapes of the ROC curves.
Looking at Figure~\ref{fig:percentages_cosine}, where the distributions of true positives computed by using the periodic and cosine kernels are compared, we see that the fraction of true positives for the cosine kernel is slightly larger than the one found through the generic periodic kernel due to the lower number of parameters. 
On the other side, the cosine kernel misses most of the sawtooth light curves, both for idealised and PTF-like baselines. 
For both kernels, the retrieved fractions for the idealised case are greater than the ones for the PTF-like light curves because of the sparser sampling of the latter.

\subsection{Parameter importance}\label{sec:param_importance}

To identify which parameters dominate the GP capability to identify periodicities, we perform a random forest regression \citep[RFR, see][]{Breiman2001}. RFR is an ensemble method able to perform regression on non-linear problems by combining the results from different decision trees, i.e. a machine learning algorithm where data are recursively split into smaller groups depending on feature (input variable) values. Each decision tree is trained using a subset of the input data and considers only a subset of the available features to introduce diversity among trees and reduce correlation. RFR is particularly useful as it allows us to gauge which feature affects the goodness of model predictions by the so-called feature importance \citep[see][]{Genuer2010}. 
In particular, to assess the feature importance, we use the permutation feature importance: a measure of how much each feature contributes to a certain classification. This is obtained by checking how much the model accuracy drops when feature values are randomly shuffled. 
If a feature is important, its shuffling leads to a large drop in predictive performance. 
\begin{table}[h!]
    \centering
    \caption{Feature importance}
    \begin{tabular}{|c|c|}
        \hline
        Feature & Feature importance \\
        \hline
        $N_{cycles}$ & 0.94 \\
        \hline
        $\sigma$ & 0.177\\
        \hline
        $A$ & 0.156\\
        \hline
        $\tau$ & 0.003\\
        \hline
    \end{tabular}
    \tablefoot{Featrure importance of the parameters of the periodic signal ($N_{cycles}$ and $A$) and of the DRW process ($\tau$ and $\sigma$)}
    \label{tab:feature_importance}
\end{table}

The feature importance of the periodic signal parameters and the ones of the DRW process are summarised in Table~\ref{tab:feature_importance}. 

\begin{figure}[t!]
    \centering
    \includegraphics[width=\linewidth]{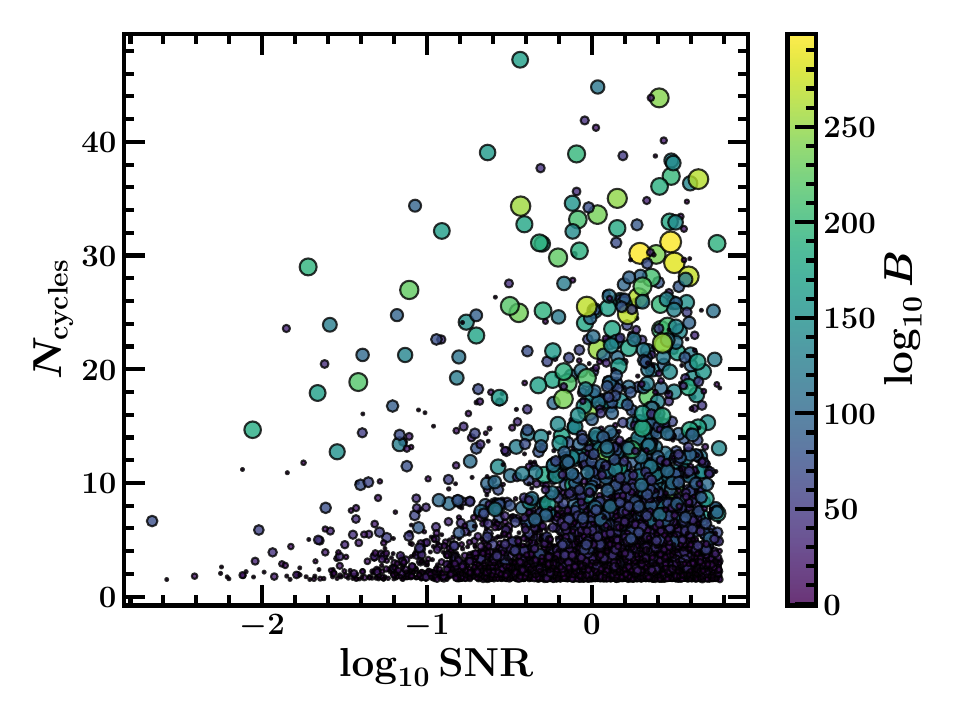}
    \caption{Number of observed cycles plotted against SNR ($A/\sigma$), colour-coded by the $\log_{10}B$. The marker size is proportional to $\log_{10}B$.}
    \label{fig:N_cycles}
\end{figure}

The number of observed cycles $N_{\rm{cycles}}=T_{\rm{obs}}/P$ is the parameter that affects the test ability to identify periodic signals the most: the higher the number of cycles observed, the better the test can identify periodicities. The periodic signal and noise amplitudes ($A$ and $\sigma$, respectively) have a relatively smaller impact on periodicity identification. 
This can be seen in Figure~\ref{fig:N_cycles}, where the light curves with a higher Bayes factor are also those with the higher number of cycles. 
From Figure~\ref{fig:N_cycles}, one can also see that light curves yielding high B tend to be found at higher SNR, with some scattering also at lower SNR. This behaviour is expected as the amplitude has the highest feature importance after the period and length of the light curve.

\subsection{Comparison with LSP based searches} \label{sec:LSP_comparison}

To define which light curves are correctly identified as periodic \cite{Lin2025} generated 100000 light curves with the same noise parameters and without periodicity for each of the 12400 light curves analysed with injected periodicity. A light curve is considered periodic if its LSP shows at least one peak whose power exceeds that of all peaks found in the noise-only simulations. To compare our results with those found through the LSP analysis, we search for the Bayes factor threshold with which we find one false positive every $10^5/25$ light curves, where the factor 25 comes from the fact that they analysed 25 frequency bins at the same time. Since we sampled only $10^4$ light curves for the combination of baselines taken into account in Figure~\ref{fig:threshold}, we use the fit shown as solid lines in the same figures to extrapolate the expected threshold for the LSST light curves, as we performed a preliminary analysis over $\sim 1000 $ light curves because of the computational cost. These values are shown in Figures~\ref{fig:threshold} and~\ref{fig:thresholds_cosine} as vertical lines. For the comparison with the LSP analysis, we fix a conservative Bayes factor threshold at $\log_{10}B_{\rm{trs}}=5$ for the PTF and ideal baselines and a threshold of $\log_{10}B_{\rm{trs}}=3$ for the LSST baselines. We note that these thresholds are much more conservative than those obtained from the ROC curves, mainly because we impose a much lower FAP in this case.

The results obtained assuming the thresholds described above are summarised in Table~\ref{tab:fractions}. In the idealised case, our analysis correctly identifies about 53$\%$ of the injected sinusoidal periodicities, $25\%$ more than those identified through the LSP analysis by \cite{Lin2025}.\footnote{To obtain the same detection fraction, we would need to raise our detection threshold to $\log(B_{\rm trs})=18.12$.} The difference is due to the nature of the algorithm, as the test statistic for periodicity detection (i.e. the Bayes factor) explicitly includes the red noise  \citep[see][]{robnik2024},
while in \citealt{Lin2025}, the test statistic is the periodogram peak for which the underlying assumption is white noise.

In the idealised sawtooth case, the difference is much more pronounced, recovering more than half of the periodicities, while \cite{Lin2025} only identified about 1$\%$ of them. While the LSP analysis significantly underperforms when searching for non-sinusoidal signals, our analysis retrieves more sawtooth signals than sinusoidal ones. This is likely due to the fact that the DRW has only one characteristic timescale, and it can either mimic the long-term variability associated with the period of the sawtooth signal, or the sharp rise occurring at the beginning of each brightening, but never both. The appearance of long-term periodicities with sharp features limits the evidence of the DRW (noise-only) model, explaining the high true positive rate of our search on the idealised sawtooth light curves.

As discussed in Section~\ref{sec:recoveries}, the efficiency of our algorithm decreases as the number of observational data points drops. Assuming a log$(B_{\rm thrs}) = 5$, only $\approx 23\%$ ($\approx 9\%$) of PTF-like lightcurves are identified as periodic in the sinusoidal (sawtooth) case. 

The LSP analysis discussed in \cite{Lin2025} seems not to suffer from the lower number of data points and the $\sim$ month-long observational gaps, actually improving their detection fraction to $\sim 39 \%$ for the PTF-like sinusoidal light curves with respect to their idealised counterparts that show a detection fraction of $\sim 28\%$. An even larger improvement (increasing from $\sim 1\%$ in the idealised baseline to $\sim 7\%$ in the PTF-like baseline) is observed in the light curves with a sawtooth periodicity. Such an increase in the LSP efficiency for sparser data is somewhat surprising  \citep[an opinion expressed in][too, see their discussion section]{Lin2025}, and might be associated with a lower degree of power leakage to frequencies associated with the one of the even sampling, see \cite{Lin2025} for more details.

\begin{table}[h!]
    \centering
    \caption{Fractions of retrieved periodicities using the thresholds from the LSP FAP}
    \begin{tabular}{|c|c|c|}
        \hline
         & Sinusoid & Sawtooth \\
         \hline
        idealised & 53.3$\%$ & 58.1$\%$ \\
         \hline
        PTF-like & 23.4$\%$ & 8,9$\%$ \\
         \hline
        LSST-like & 50.9$\%$ & 64.5$\%$ \\
         \hline
    \end{tabular}
    \tablefoot{Fraction of light curves that show a Bayes factor greater than the threshold of $\log_{10}B_{\rm{trs}}=5$ for the ideal and PTF baseline and $\log_{10}B_{\rm{trs}}=3$ for the LSST baseline.}
    \label{tab:fractions}
\end{table}

By fixing the Bayes factor threshold to $\log_{10}B_{\rm{trs}}=5$ for the idealised and PTF baselines or $\log_{10}B_{\rm{trs}}=3$ for the LSST baseline, the accuracy of the period recovery improves. 
The fraction of light curves showing a retrieved period within $15\%$ of the injected one among all the light curves with a Bayes factor greater than the threshold for all the combinations of baselines and periodicity shapes, is summarised in Table~\ref{tab:goodness_comparison}.
\begin{table}[h!]
    \centering
    \caption{Fractions light curves with a retrieved period within $15\%$ of the injected one using the thresholds from the LSP FAP}
    \begin{tabular}{|c|c|c|}
        \hline
         & Sinusoid & Sawtooth \\
         \hline
        idealised & 92.2$\%$ & 99.8$\%$ \\
         \hline
        PTF-like & 87.2$\%$ & 91.1$\%$ \\
         \hline
        LSST-like & 96.1$\%$ & 98.9$\%$ \\
         \hline
    \end{tabular}
    \tablefoot{Fraction of light curves that show a retrieved period within $15\%$ of the injected one among the light curves with a Bayes factor greater than the threshold of $\log_{10}B_{\rm{trs}}=5$ for the ideal and PTF baseline and $\log_{10}B_{\rm{trs}}=3$ for the LSST baseline.}
    \label{tab:goodness_comparison}
\end{table}

\begin{figure*}
    \centering
    \includegraphics[width=\linewidth]{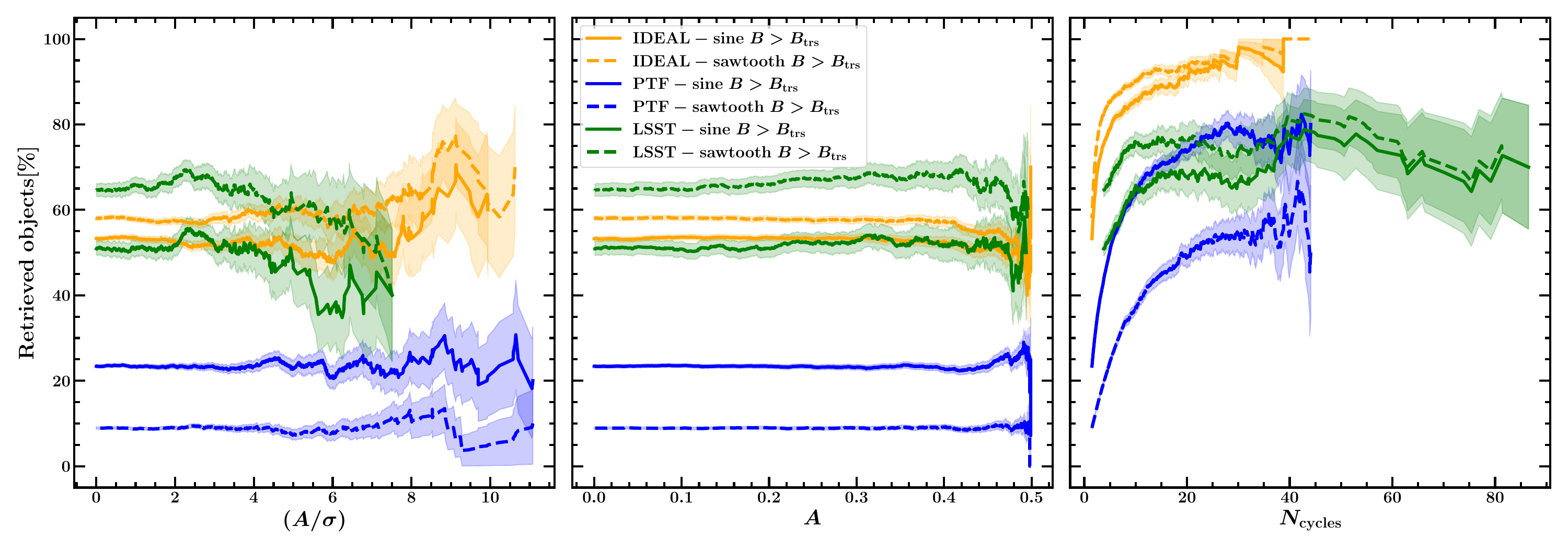}
    \caption{Fraction of retrieved periodicities as a function of signal-to-noise ratio (SNR), periodic signal amplitude ($A$), and number of observed cycles ($N_{\rm{cycles}}$), shown from left to right. The shaded regions refer to the uncertainties on the points.}
    \label{fig:function_of_parameters}
\end{figure*}

Finally, in Figure~\ref{fig:function_of_parameters}, we report the fraction of retrieved periodic signals as a function of the parameters as reported by \cite{Lin2025}, where instead of the period we plot the number of cycles, as we identified it as the most important parameter for our analysis, see Section~\ref{sec:param_importance}. 
To do so, we divide the range of the injected SNR ($A/\sigma$), amplitude ($A$) and the number of observed cycles ($N_{\rm{cycles}}$) into 10 bins and compute the fraction of retrieved periodic light curves among all the light curves in each bin.  
We find, as expected from the previous discussion, a decrease in the fraction of retrieved light curves as the period of the signal increases.  

\section{Discussion and conclusions}\label{sec:conclusions}

In this work, we propose a Bayesian framework based on GPs for the identification of
arbitrarily shaped periodicities in evenly and unevenly sampled light curves. The method leverages Gaussian processes to model both periodic and stochastic variability and on nested sampling to determine the relative evidence of a periodicity in the data and to constrain the associated periods. 

We studied the performance of the method characterising the fractions of true positive (periodic signals correctly identified as periodic) and of false positive (realisations of standard AGN variability misidentified as periodic signals) as a function of the Bayes factor $B$, defined as the ratio between the evidence of a model including 
a periodic component and a model with only non-periodic variability. An optimal threshold $B_{\rm trs}$ has been identified for each type of light curve (characterised by the duration of the monitoring campaign $T_{\rm obs}$ and the frequency and regularity of the observations) to maximise the fraction of true positive detections while limiting the contamination by false positives. 
With such threshold the percentages of true positives are as high as $\approx 70\%$ for idealized baselines (even sampling every day with durations extracted from real PTF lightcurves) independently of the periodicity shape, and decrease down to $\approx 53\%$ ($\approx 26\%$) for sinusoidal (sawtooth) periodicities sampled with the same $T_{\rm obs}$ of the idealized ones and with uneven sampling of real PTF data (PTF baseline). The decreased efficiency of the algorithm for sparser data and sawtooth-shaped signals is associated with the possibility that the rapid rise of luminosity can fall in data gaps, hindering the correct identification of periodicities. For the same reason, while in the other cases $>80\%$ of the periods retrieved are accurate within a relative error of $15\%$, PTF baselines recover the wrong period in up to $\sim 40 \%$ of the light curves identified as periodic.
In all the considered scenarios, the false alarm probability is $<15 \%$
\footnote{Once selected, the FAP of each light curve can be individually evaluated from its value of $B$.}.

The generic periodic GP kernel adopted here is more versatile than the cosine kernel previously used in literature for the analysis of AGN light-curves \citep{Zhu_2020, foustoul25}: while the cosine kernel has comparable performances for sinusoidal signals recovers $<6\%$ of sawtooth periodicities even after the optimisation of the $B_{\rm trs}$. Such a low fraction of true positives is not unexpected, as the cosine kernel struggles in modelling at the same time the correct period and sharp features in the light curves.

We completed our comparison study by confronting the performances of our model with the study by \cite{Lin2025} based on Lomb-Scargle-periodogram analyses. These searches identify periodic signals on the basis of a false alarm threshold. We therefore identified a conservative threshold to be used to perform a fair comparison with the results presented in \cite{Lin2025} at $\log_{10}B=5$ for idealised and PTF-like light curves and $\log_{10}B=3$ for LSST-like light curves.
With this new (significantly higher) threshold, the fraction of identified modulations in idealised sinusoidal light curves decreases to $\sim 53\%$,  higher than the $\sim 28 \%$ obtained  
using the LSP-analysis. Even more relevant is the increase of recovered periodicities in the idealised sawtooth case, with more than $58\%$ of the systems being identified compared to the $\sim 1\%$ recoveries in \cite{Lin2025}.
As for the ROC-optimised $B_{\rm trs}$, even assuming $\log_{10} B_{\rm trs}=5$, we find that periodicities in PTF-like light curves are retrieved in a much smaller fraction when compared to the idealised case because of their sparse sampling. Surprisingly, such a trend is not observed in \cite{Lin2025}, who find a significant increase (up to a factor of $\approx 7$ for sawtooth light curves) in the recovery fraction with less and sparser data with quasi-periodic gaps. A more detailed comparison to clarify the reason for such behaviour is deferred to a future investigation.  

The main feature determining the recovery fraction of periodic signals in our model is the number of periods present in the light curves. To gauge the performance of our method on longer surveys and to test its relevance for the starting LSST campaign, we selected the properties (both of the periodic component and of the noise) of 1000 light curves that were not identified with PTF baselines and resampled them with LSST-like baselines. Of these 1000, about $60 \% \ (70\%)$ light curves are correctly identified as periodic with a small fraction of false positives, around $\sim2\% \ (3\%)$ in the sinusoidal (sawtooth) case, respectively.  

The ability to identify non-sinusoidal shapes of periodic signals is of fundamental importance for the search for MBHBs. Indeed, saw-tooth-like light curves are expected when the periodicity is imprinted by the periodic feeding of the two massive black hole minidiscs (caused by the periodic non-axisymmetric potential of the binary). Deviations from sinusoidal modulations are also expected for periodicities caused by periodic gravitational lensing involving an MBHB or by Doppler boosting in an eccentric binary, even in the idealised scenario of constant intrinsic luminosity. 

In principle our analysis can be tailored for the identification of non-active MBHB, if a luminous star in the binary host galaxy is periodically lensed by the two MBHs \citep{lensing25}, and can be used for astrophysical systems other than MBHBs, such as quasi-periodic eruptions \citep[QPE, see][]{Miniutti_2019, Giustini_2020, Arcodia_2021}, i.e. high amplitude bursts of X-ray radiation recurring every few hours and originating near the central supermassive black holes of galactic nuclei, 
or even planetary transits used for the search of exoplanets by looking for periodic dips in the flux of a star \citep[see][]{Holman2005, winn2010}. In Appendix~\ref{app:spiky} we demonstrate that the test can retrieve the correct periodicities also in the case of sharp and narrow periodic modulations in the light curves, as expected in the examples above. In this case, we find that, unlike the sinusoidal and sawtooth cases, the signal-to-noise ratio plays a crucial role in the detection of periodicities. This is because, for spiky periodicities, most of the signal is concentrated within narrow time windows and is completely absent outside them. As a result, once the noise amplitude becomes comparable to that of the spiky signal, our analysis can no longer reliably distinguish the periodic component from the noise.

In Appendix~\ref{app:time_comparison}, we compare the computational cost of the algorithm of this work with that of the Lomb-Scargle analysis as implemented in \cite{Charisi2016}, showing that the algorithm presented here is more computationally expensive. We stress the fact that, along with this additional computational cost, comes the advantage of a full Bayesian framework. This provides posterior distributions on the model parameters and allows for a robust model comparison via the Bayesian evidence, which is not available with the Lomb-Scargle periodogram.

We identified two main aspects that can be improved in our search strategy. On the one hand, it is essential to simultaneously leverage photometric points collected in multiple bands.\footnote{For example, the Wide-Fast-Deep survey mode of LSST \citep[comprising roughly 90\% of the total survey time, see][]{LSST} will achieve an average sampling rate of one observation every $\sim 3$ days only when all the photometric bands are considered.}.

We plan to improve our analysis using multi-output Gaussian processes to allow for the parallel analysis of all the available bands so that they can inform one another. Such a procedure has a high computational cost, as the numerical burden scales approximately as the cube of the number of observations \footnote{We notice that the effective computational cost of the likelihood evaluation, which dominates the computational burden of the algorithm, strongly depends on the implementation and on the kind of covariance matrix assumed. Examples of different scalings are \texttt{GpyTorch} \citep[see][and Appendix~\ref{app:time_comparison}]{gpytorch} and \texttt{celerite} \citep[see][]{celerite}} in all the bands. This is a major limitation of our method, which is still too slow to be applied to the entire LSST quasar sample. For this reason, we are working on the parallelisation of our algorithm on GPUs.

This method can be further generalised by considering other noise models, e.g. the Damped Harmonic Oscillator \citep[DHO, see][]{Yu_2022}. Such an analysis is not performed in this first exploratory work since the light curves we used to compare our performances with alternative searches were generated assuming the DRW as the noise model. The characterisation of the performances of our search for different noise models is deferred to a future investigation.

\begin{acknowledgements}
LB acknowledges ISCRA for awarding this project access to the LEONARDO supercomputer, owned by the EuroHPC Joint Undertaking, hosted by CINECA (Italy). LB wishes to thank the "Summer School for Astrostatistics in Crete" for providing training on the statistical methods adopted in this work.
MD and RB acknowledge support from the ICSC National Research Center funded by NextGenerationEU, and financial support by the Italian Space Agency grant Phase B2/C activity for LISA mission, Agreement n.2024-NAZ-0102/PE. MC is funded by the European Union (ERC-StG-2023, MMMonsters, 101117624). JCR acknowledges support from the National Science Foundation (NSF) from grant NSF AST-2205719 and the NASA Preparatory Science program under award 20-LPS20-0013.

\end{acknowledgements}

\bibliographystyle{aa} 
\bibliography{main}
\begin{appendix}
\section{The periodic kernel}\label{app:spec_periodic}
\begin{figure}[h!]
        \centering
        \includegraphics[width=0.8 \linewidth]{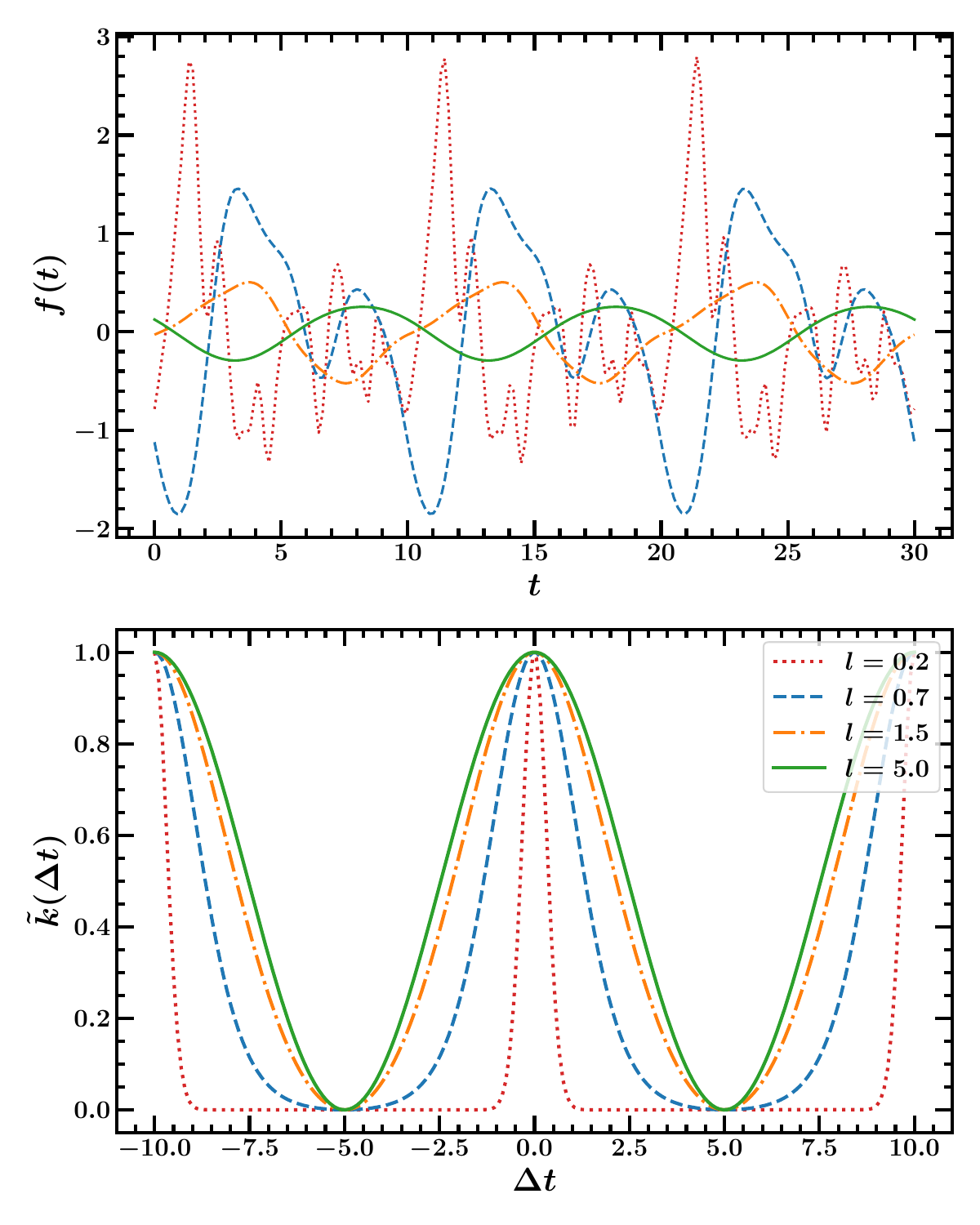}
        \caption{Upper panel: realisations of a Gaussian process with a periodic kernel at different lengthscales.  Lower panel: rescaled periodic kernel $\tilde{k}_{\rm periodic} (\Delta t)$, defined in Equation~\ref{eq:rescaled_kernel}, with different lengthscales. Lengthscales, colours and linestyles are shared between the two panels.}
        \label{fig:periodic_kernel_different_l}
    \end{figure}

In Section~\ref{sec:periodic_kernel}, we briefly discussed how the periodic kernel is defined and how it describes signals with arbitrary shapes thanks to the presence of a kernel lengthscale. 
In the top panel of Figure~\ref{fig:periodic_kernel_different_l}, we show how realisations of the Gaussian process change when changing the kernel lengthscale. In the bottom panel, we show, for visualisation purposes, the rescaled periodic kernel defined as 
\begin{equation}\label{eq:rescaled_kernel}
    \tilde{k}(\Delta t)=\frac{k_{\rm{periodic}}(\Delta t) - \rm{min}(k_{\rm{periodic}}(\Delta t))}{\rm{max}(k_{\rm{periodic}}(\Delta t))-\rm{min}(k_{\rm{periodic}}(\Delta t))},
\end{equation}
where $\Delta t =|t_i-t_j|$ with $t_i,t_j$ being two observations of the timeseries. 
Note that changes when changing its lengthscale. 
From Figure~\ref{fig:periodic_kernel_different_l}, it is possible to see that, as the lengthscale decreases, only observations that are very close in time or separated by an integer multiple of the period remain strongly correlated, while the correlation drops rapidly for points in between these peaks, allowing for the modelling of sharp features in the light curves.
    
The flexibility of the periodic kernel can also be explained 
by means of the Fourier-Bessel expansion. The periodic kernel can be rewritten as:
\begin{equation}
        k_{\rm{periodic}}(t_i, t_j) = A^2 \exp\left(-\frac{1}{l^2}\right) \exp\left[\frac{1}{l^2}\cos\left(\frac{2\pi}{P} | t_i - t_j | \right)\right] .
    \label{eq:periodic_kernel_expanded}
    \end{equation}
    By recalling that $\exp[a\cos(\theta)]$ can be expanded as:
        \begin{equation}
        \exp[a\cos(\theta)] = I_0(a) + 2\sum_{n=1}^\infty I_n(a) \cos(n\theta),
    \end{equation}
    where $I_n$ are the modified Bessel functions, we get:
    \begin{equation}
\begin{split}
k_{\rm periodic}(t_i, t_j) &= A^2 \exp\left(-\frac{1}{l^2}\right) \Bigg[ 
 I_0\left(\frac{1}{l^2}\right) \\
 &+ 2\sum_{n=1}^\infty I_n\left(\frac{1}{l^2}\right) 
\cos\left(\frac{2\pi n}{P} |t_i - t_j|\right)
\Bigg].
\end{split}
\end{equation}

From this expansion, it becomes clear that the kernel describes an infinite series of Fourier modes with weights $\exp(-1/l^2) I_n(1/l^2)$. As $l^2 \rightarrow \infty$, the only significant term is the $n=1$ harmonic as the Bessel weights $I_n(1/l^2)$ vanish for $n>1$ when the argument is small, while for $l^2 \rightarrow 0$, many harmonics contribute to the signal. Such an expansion shows that, similarly to the Lomb–Scargle periodogram, the periodic kernel represents a periodic signal as a decomposition into harmonics. However, instead of explicitly fitting independent amplitudes for each Fourier mode, the relative weighting of the harmonics is implicitly controlled by the kernel lengthscale. This allows the periodicity search algorithm to self-consistently leverage the power at different harmonics, resulting in a higher retrieved fraction compared to the LSP-based algorithm searching for an excess of signal at only one frequency.

\section{Searching for spiky signals}\label{app:spiky}
\begin{figure}
    \centering
    \includegraphics[width=0.8 \linewidth]{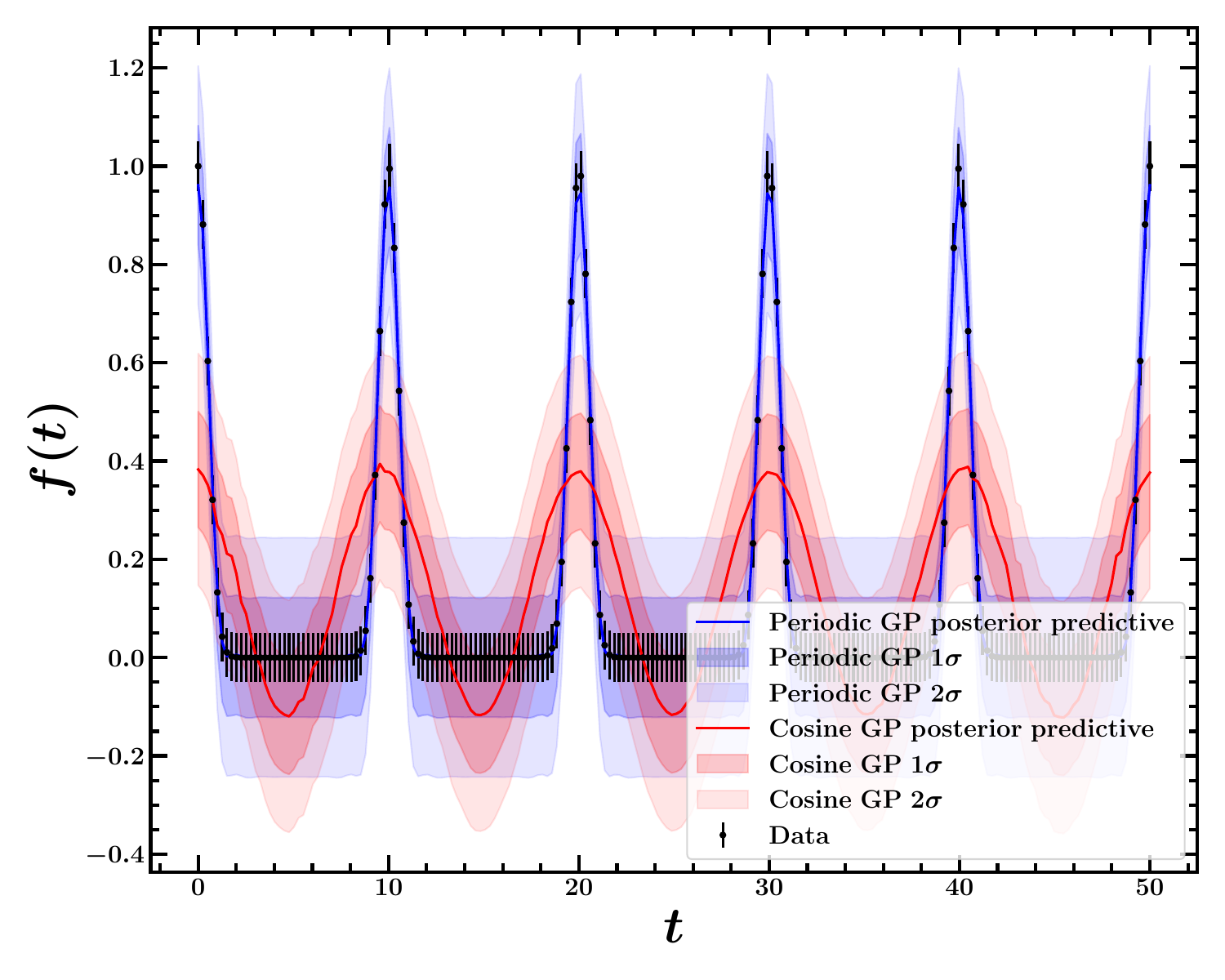}
    \caption{Posterior predictive distributions obtained by fitting a "spiky" light curve using the cosine kernel (red solid line and shaded region) and periodic kernel (blue solid line and shaded region) defined in Equations~\ref{eq:cos_kernel} and~\ref{eq:periodic_kernel}, respectively. The light curve is sampled 4 times per day. Similarly to the sawtooth case in Figure~\ref{fig:fit_examples}, the cosine kernel is unable to properly describe the shape of the periodic signal, while the periodic kernel manages to capture it. The time $t$ and signal $f(t)$ are reported in arbitrary units.}
    \label{fig:spike_even}
\end{figure}

\begin{figure}
    \centering
    \includegraphics[width=0.8 \linewidth]{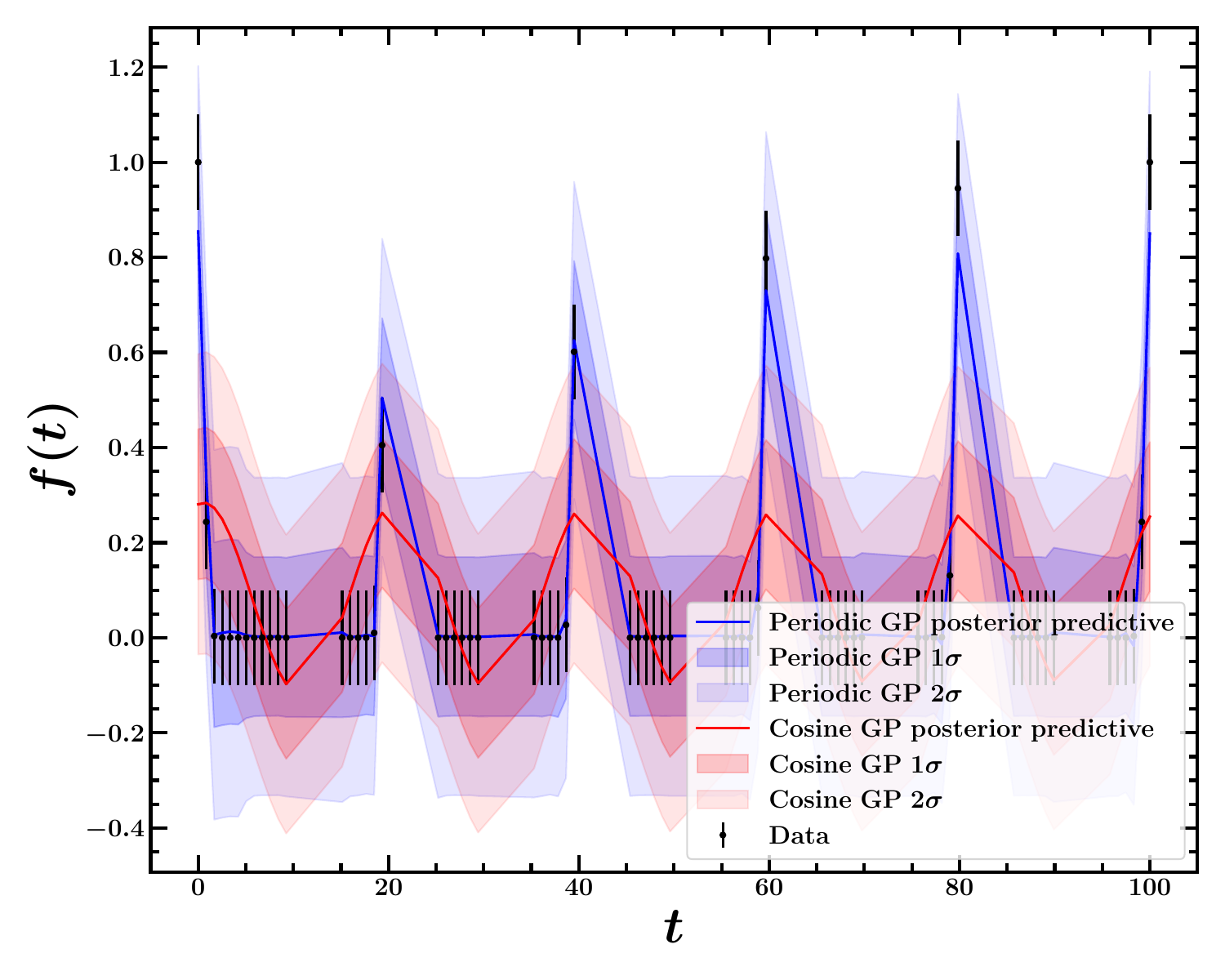}
    \caption{Posterior predictive distributions obtained by fitting a "spiky" light curve using the cosine kernel (red solid line and shaded region) and periodic kernel (blue solid line and shaded region) defined in Equations~\ref{eq:cos_kernel} and~\ref{eq:periodic_kernel}, respectively. The light curve is made sparser with respect to the light curve shown in Figure~\ref{fig:spike_even} by the introduction of gaps in the data. The time $t$ and signal $f(t)$ are reported in arbitrary units. }
    \label{fig:spike_uneven}
\end{figure}

As briefly mentioned in Section~\ref{sec:introduction}, in both cases of active or dormant massive black holes, depending on the inclination of the system, "spiky" signals are expected because of gravitational lensing, either the lensing of one massive black hole on its companion or because of the lensing of the MBHB onto a background star \citep[see][]{lensing25}. Similarly to the case of sawtooth signals, when computing the periodogram, the power leaks to frequencies different from the inverse of the true period. Thus, the LSP analysis would encounter the same problems as in the sawtooth case, while the GP analysis would also be able to detect this kind of signal. In Figure~\ref{fig:spike_even}, we show that, in the absence of red noise and with an idealised baseline characterised by an even sampling and a cadence sufficient to resolve the sudden rise and fall of the flux (the light curve is sampled 4 times per day in this case), the GP analysis can successfully identify both the shape and period of spiky signals when using the periodic kernel. Similarly to the sawtooth case in Figure~\ref{fig:fit_examples}, the cosine kernel is unable to properly describe the shape of the spiky signal. For the light curve shown in Figure~\ref{fig:spike_even}, the Bayes factor between the periodic and cosine kernel is $\log_{10} E_{\mathrm{periodic}}/E_{\rm{cosine}} \sim 300$. When windows in the sampling are introduced, the GP test begins to struggle and starts to identify longer periods in addition to the true one. This effect can be seen in Figure~\ref{fig:spike_uneven}, where one can see that multimodalities in the retrieved period are introduced due to the windowing. Also in this case, though, the cosine kernel seems unable to describe the spiky nature of the signal, with the periodic kernel being favoured with a Bayes factor of  $\log_{10} E_{\mathrm{periodic}}/E_{\rm{cosine}} \sim 200$.

\begin{figure}
    \centering
    \includegraphics[width=0.8 \linewidth]{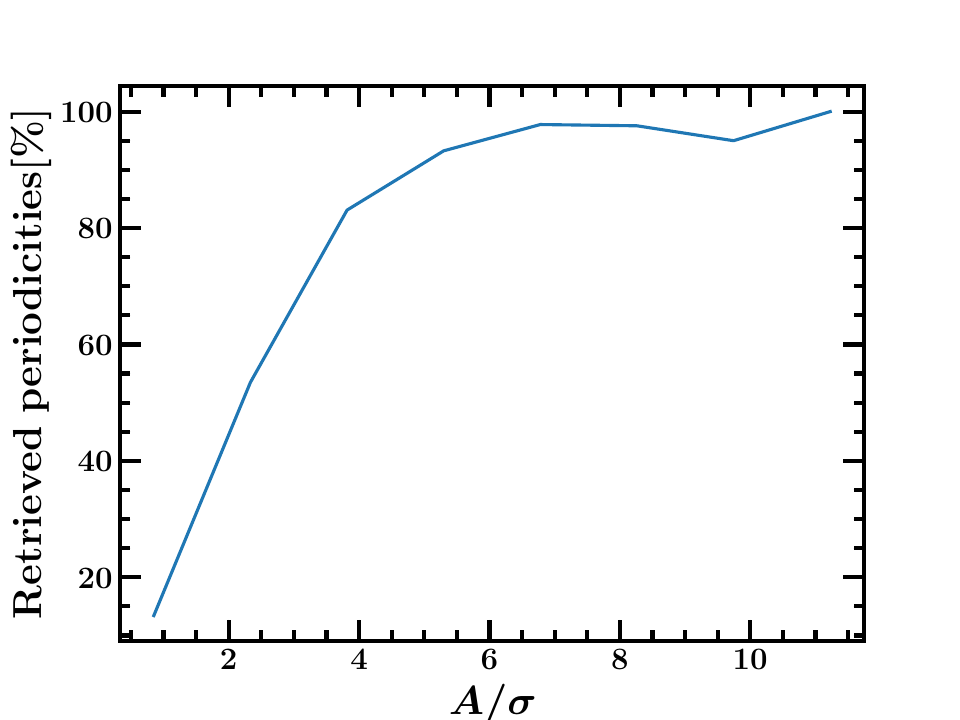}
    \caption{Fraction of retrieved periodicities as a function of signal-to-noise ratio (SNR)}
    \label{fig:spike_SNR}
\end{figure}

We repeated the same study done for the sinusoidal and sawtooth light curves for 3000 evenly sampled spiky light curves, finding that $42.53\%$ of the periodicities are detected. The main difference from the previous results is that spiky light curves show a strong dependence on the signal-to-noise ratio, as shown in Figure~\ref{fig:spike_SNR}. This arises because the periodic signal is only visible for a very brief fraction of the observation; during the rest of the time, the light curve is dominated by noise. Consequently, when the noise amplitude becomes comparable to that of the periodic signal, the signal is effectively lost, unlike the sinusoidal or sawtooth cases, where the signal is present more continuously.

\section{Computational cost comparison}\label{app:time_comparison}
As mentioned throughout this paper, the algorithm we propose, which leverages Gaussian and nested sampling, scales very differently compared to the more common Lomb-Scargle periodogram analysis. The latter algorithm, as implemented by \cite{Charisi2016} and \cite{Lin2025}, with which we are comparing our results, is dominated by the generation of DRW light curves to assess the probability for a peak observed in the real light curve to be explained by pure-noise realisations. The algorithm presented in this work is instead dominated by the number of likelihood evaluations necessary for the nested sampler to converge, and the computational cost of each likelihood evaluation. 
\begin{figure} [h!]
    \centering
    \includegraphics[width=0.83 \linewidth]{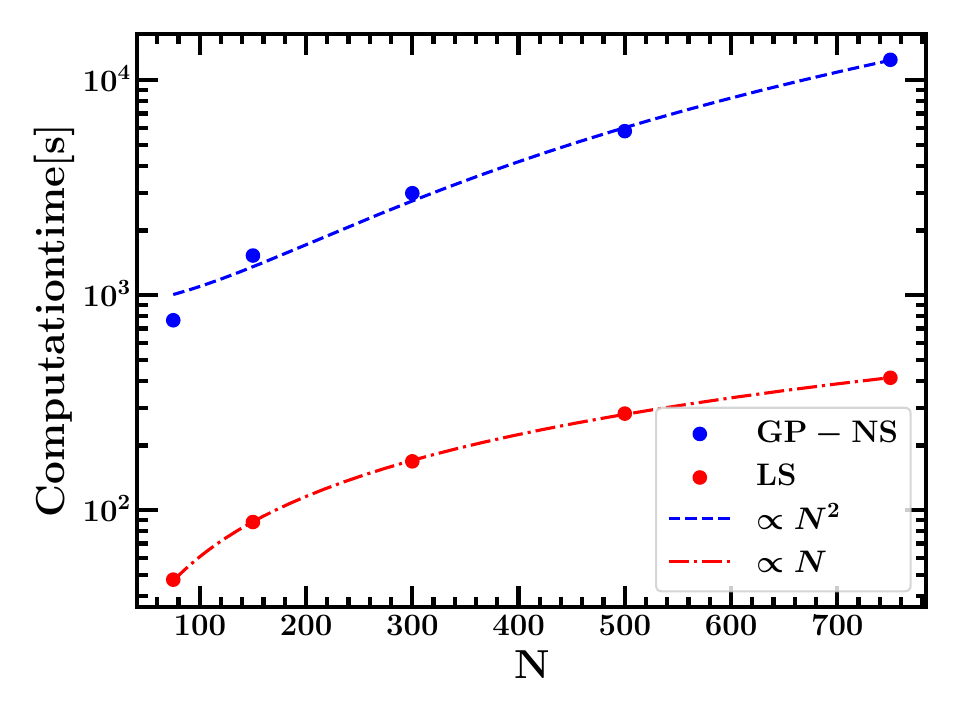}
    \caption{Comparison between the computational cost in seconds as a function of the number of points in an idealised light curve for the algorithm proposed in this work (blue points) and the Lomb-Scargle analysis as implemented in \cite{Charisi2016} (red points). The blue and red dashed lines show a power law fit to the data.}
    \label{fig:timing_comparison}
\end{figure}
In Figure~\ref{fig:timing_comparison}, we show the very different scaling of the computational times of the two algorithms.
We fitted the times to assess the scaling with the number of points, observing that the Lomb-Scargle analysis scales roughly linearly with the number of points, while the algorithm presented in this work scales roughly as $N^2$, where $N$ is the number of points in the light curve.  We notice that the computational cost, measured as end-to-end runtimes, does not scale as the computational cost of a GP with an $N\times N$ covariance matrix. This is due to the way in which the likelihood is evaluated by \texttt{GPyTorch}, which uses a batched version of linear conjugate gradients \citep[see][]{gpytorch} instead of a Cholesky decomposition. This reduces the time complexity of exact GP likelihood evaluation from $O(N^3)$ to $O(N^2)$. We also note that the total runtime captures any variation in the number of likelihood evaluations required by the nested sampler to reach convergence, which may itself depend on the complexity of the posterior and the data quality.  
The light curves used to evaluate the computational cost are idealised, and both the Lomb-Scargle and GP analyses have been run with no further parallelisation.

\end{appendix}
\end{document}